\documentclass{article}

\usepackage[margin=1in]{geometry}
\usepackage[numbers,sort&compress]{natbib}
\usepackage{amsmath}
\usepackage{mathtools}
\usepackage{MnSymbol}
\usepackage{graphicx}
\usepackage{bbm}

\DeclareMathOperator*{\CP}{\Box}
\DeclareMathOperator*{\DP}{\times}
\DeclareMathOperator*{\SP}{\boxtimes}
\DeclareMathOperator*{\KS}{\oplus}
\DeclareMathOperator*{\KP}{\otimes}
\DeclareMathOperator*{\GP}{\boxtimes}
\DeclareMathOperator*{\KG}{\mathrlap{\oplus}\otimes}

\newcommand{\bone}{\mathbbm{1}}

\begin{document}

\title{Graph Product Multilayer Networks: Spectral Properties and Applications}

\author{Hiroki Sayama$^{1,2,3,4}$\\
$^1$ Center for Collective Dynamics of Complex Systems,\\
Binghamton University, Binghamton, New York 13902, USA\\
$^2$ Max Planck Institute for the Physics of Complex Systems,\\
D-01187 Dresden, Germany\\
$^3$ Center for Complex Network Research and Department of Physics,\\
Northeastern University, Boston, Massachusetts 02115, USA\\
$^4$ Faculty of Commerce, Waseda University, Shinjuku, Tokyo 169-8050, Japan}

\date{}

\maketitle

\begin{abstract}
This paper aims to establish theoretical foundations of graph product
multilayer networks (GPMNs), a family of multilayer networks that can
be obtained as a graph product of two or more factor networks.
Cartesian, direct (tensor), and strong product operators are
considered, and then generalized. We first describe mathematical
relationships between GPMNs and their factor networks regarding their
degree/strength, adjacency, and Laplacian spectra, and then show that
those relationships can still hold for nonsimple and generalized
GPMNs. Applications of GPMNs are discussed in three areas: predicting
epidemic thresholds, modeling propagation in nontrivial space and
time, and analyzing higher-order properties of self-similar
networks. Directions of future research are also discussed.\\
Keywords: graph product, multilayer networks, degree/adjacency/Laplacian spectra, epidemic thresholds, propagation, self-similar networks
\end{abstract}

\section{Introduction}
\label{sec:intro}

Multilayer networks have gained a lot of attention in the network
science and complex systems science communities over the last several
years \cite{kivela2014multilayer,boccaletti2014structure,dedomenico2016}. Multilayer
networks describe complex systems in multiple subsystems (layers) and
their interconnectivities. This offers an intuitive, powerful, and
practical framework for modeling and analysis of complex systems.

In the multilayer networks literature, {\em graph product}, a
multiplication of two or more graphs discussed in discrete
mathematics, is often used as a representation of multilayer network
topology
\cite{leskovec2010kronecker,sole2013spectral,de2013mathematical,asllani2015turing,brechtel2016master}. It
is sometimes used explicitly and visually (e.g., Cartesian product of
two graphs), or at other times implicitly through application of
non-conventional product to matrices (e.g., Kronecker product of
adjacency matrices). To the best of our knowledge, however, there is
an apparent lack of systematic references that summarize various
properties of such graph-product-based multilayer network topologies
and their implications for the structure and dynamics of complex
networks. The present study aims to meet this need.

In this paper, we aim to establish theoretical foundations of graph
product multilayer networks (GPMNs), a family of multilayer networks
that can be obtained as a graph product of two or more factor
networks. We primarily consider the following three major graph
product operators: Cartesian, direct (tensor), and strong products. We
describe fundamental mathematical relationships between GPMNs and
their factor networks regarding their degree/strength, adjacency, and
Laplacian spectra (i.e., eigenvalues of degree/strength, adjacency,
and Laplacian matrices). These relationships are exact, except for
Laplacian ones of direct and strong products, while those Laplacian
spectra can also be approximated using heuristic methods
\cite{sayama2016estimation}. We also extend the definitions of GPMNs
to nonsimple networks with directed, weighted, signed, and/or
self-looped edges, and then generalize graph product operation to
arbitrary linear combination of Cartesian and direct products. We show
that the previously reported spectral relationships between GPMNs and
their factor networks are still maintained in nonsimple and
generalized cases. In the latter half of this paper, we demonstrate
the effectiveness of GPMNs through three applications: prediction of
epidemic thresholds, modeling of propagation in nontrivial space and
time, and analysis of higher-order properties of self-similar
networks.

It should be noted here that most real-world complex networks are not
GPMNs, and therefore their structure and dynamics would not be
fully captured within the GPMN framework. However, GPMNs may still be
used as a reference or surrogate model to which real-world multilayer
network structure and dynamics can be compared. One may test various
structural/dynamical properties of complex network models or data
against GPMN-based approximations obtained by assuming that
intra-layer networks are identical across all layers and that
inter-layer edges follow certain patterns. There are also other areas
of applications of GPMNs, as illustrated later in this paper.

The rest of the paper is structured as follows. In the next section,
we first define simple GPMNs and summarize their known spectral
properties. In Section \ref{sec:nonsimple}, we extend the definitions
of GPMNs to nonsimple graphs. In Section \ref{sec:generalized-gp}, we
generalize graph product operation. Sections \ref{sec:app1},
\ref{sec:app2}, and \ref{sec:app3} describe applications of
GPMNs. Finally, Section \ref{sec:conclusions} concludes this paper
with a discussion on directions of future research.

\section{Simple Graph Product Multilayer Networks and Their Spectral Properties}
\label{sec:simple}

We define {\em graph product multilayer networks (GPMNs)} as a
particular family of multilayer networks that can be obtained by
applying graph product operation(s) to two or more smaller
networks. We call those smaller networks {\em factor networks}. One of
the factor networks is often considered a template of intra-layer
networks that uniformly applies to all the layers, while other factor
networks are often considered to represent aspects of layers
\cite{kivela2014multilayer} that can be nonlinear and network-shaped.

\begin{figure}[tp]
\centering
\begin{tabular}{lcll}
$G:$ & \raisebox{-0.5\height}{\includegraphics[scale=0.38]{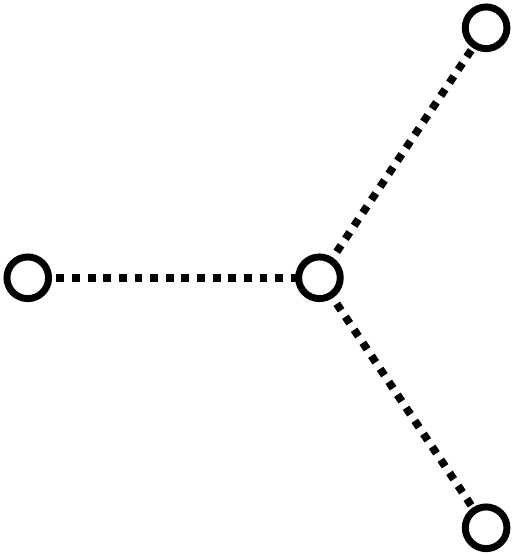}} & ~~~ &
$\displaystyle A_G = \begin{pmatrix}%
0 & 1 & 0 & 0\\
1 & 0 & 1 & 1\\
0 & 1 & 0 & 0\\
0 & 1 & 0 & 0
\end{pmatrix}$\\
&&\\
$H:$ & \raisebox{-0.4\height}{\includegraphics[scale=0.38]{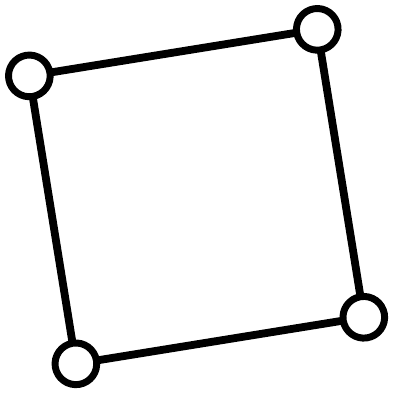}} & ~~~ &
$\displaystyle A_H = \begin{pmatrix}%
0 & 1 & 1 & 0\\
1 & 0 & 0 & 1\\
1 & 0 & 0 & 1\\
0 & 1 & 1 & 0
\end{pmatrix}$\\
&&\\
&&\\
\end{tabular}\\
\begin{tabular}{ll}
(a) Cartesian product $G \CP H$ & \\
\hspace*{0.6in}\raisebox{-0.5\height}{\includegraphics[scale=0.39]{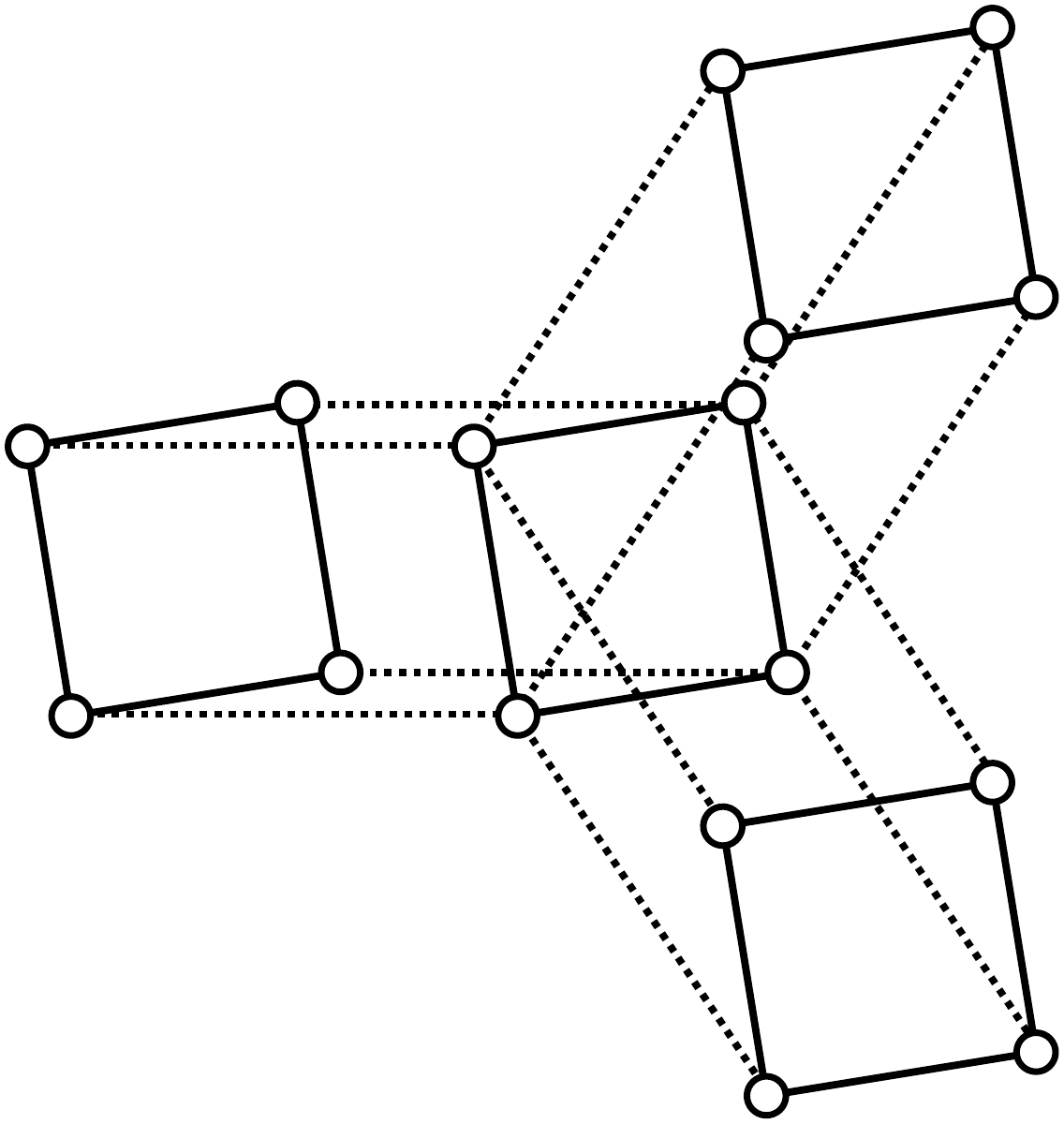}} &
\begin{minipage}{3in}
$\begin{array}{rl}%
A_{G \CP H} &= A_G \KS A_H\\
&=A_G \KP I_{|V_H|} + I_{|V_G|} \KP A_H\\
&=\left(\raisebox{-0.48\height}{\includegraphics[scale=0.29]{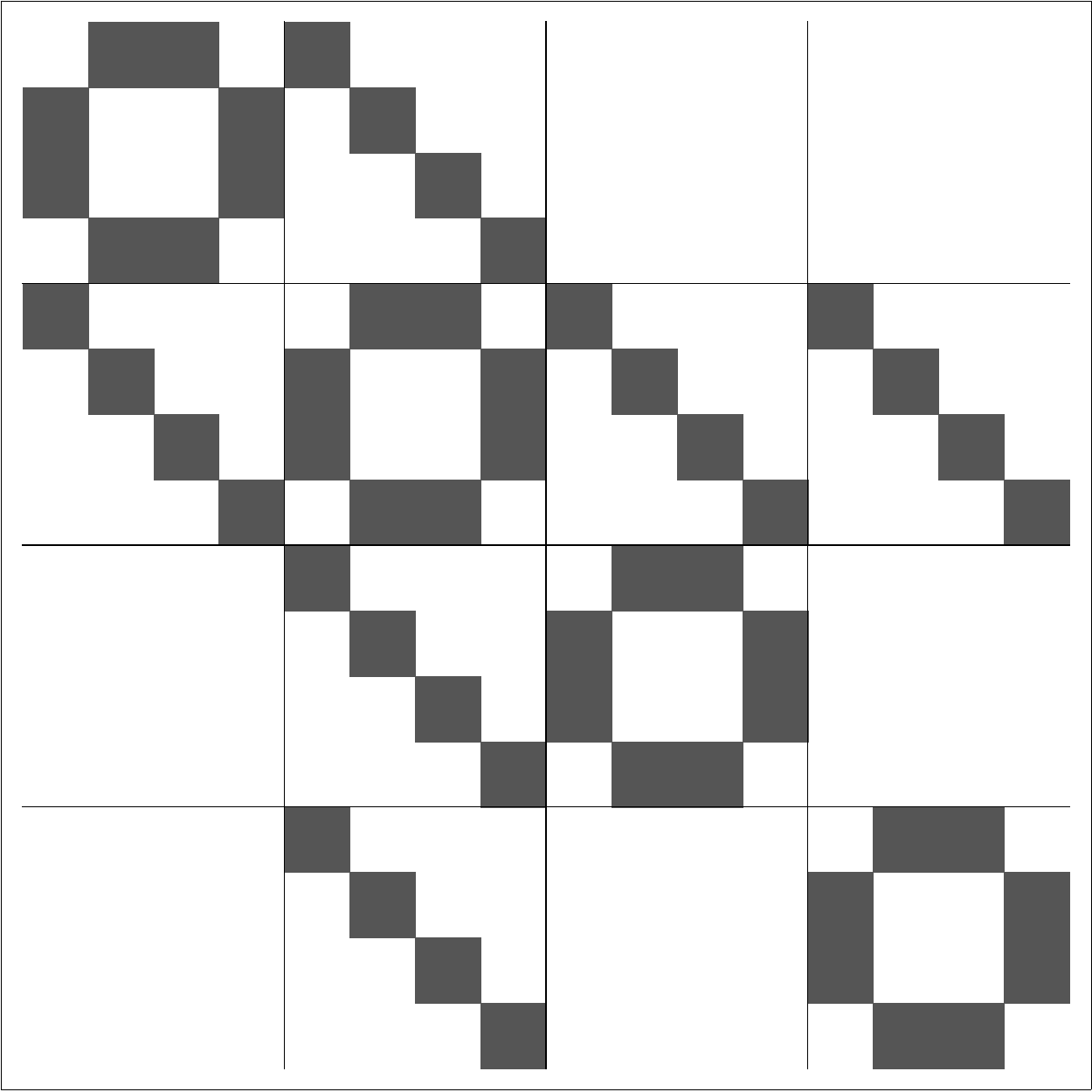}}\right)
\end{array}$
\end{minipage}\\
&\\
(b) Direct (tensor) product $G \DP H$ & \\
\hspace*{0.6in}\raisebox{-0.5\height}{\includegraphics[scale=0.39]{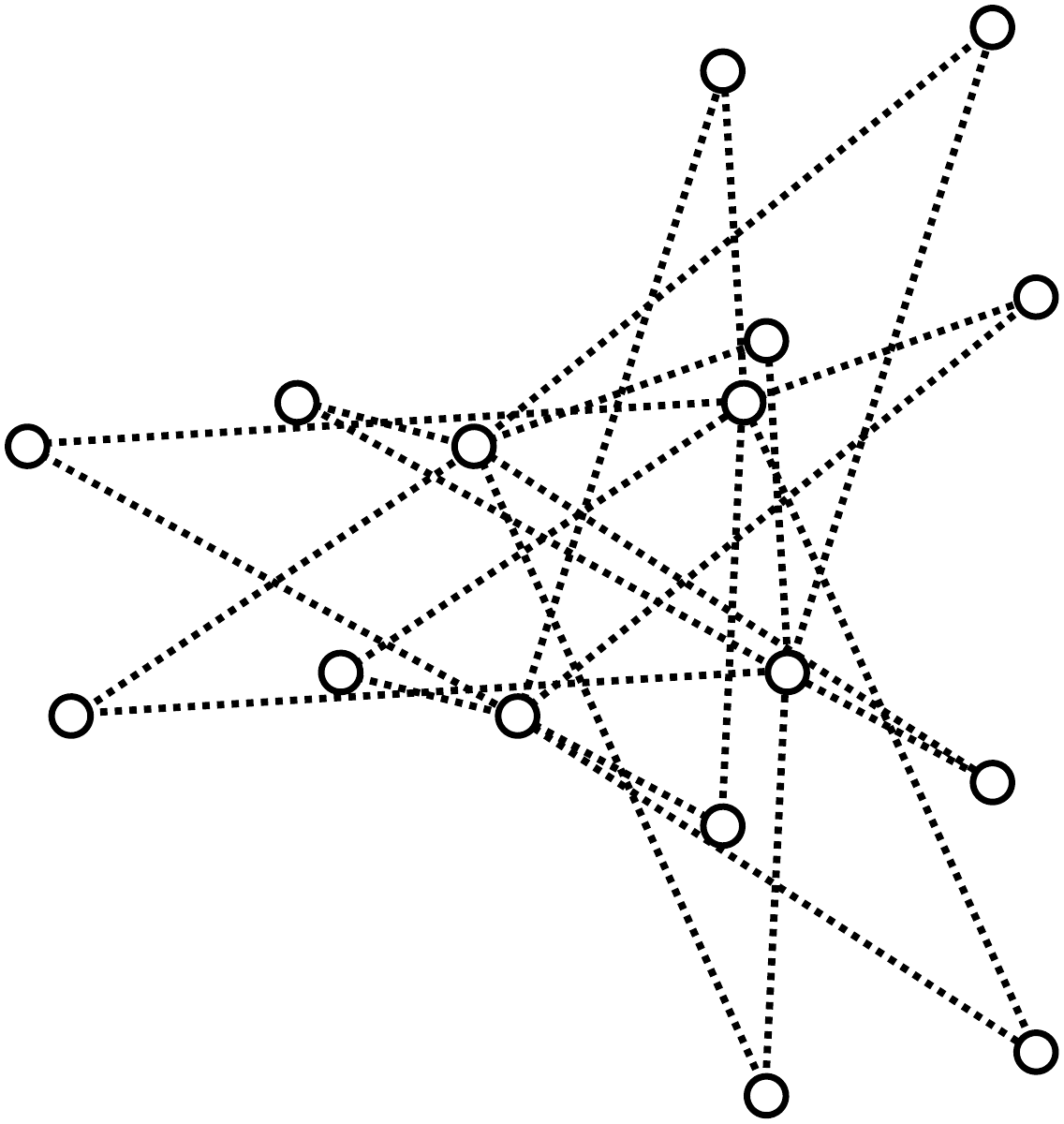}} &
\begin{minipage}{3in}
$\begin{array}{rl}%
A_{G \DP H} &= A_G \KP A_H\\
&=\left(\raisebox{-0.48\height}{\includegraphics[scale=0.29]{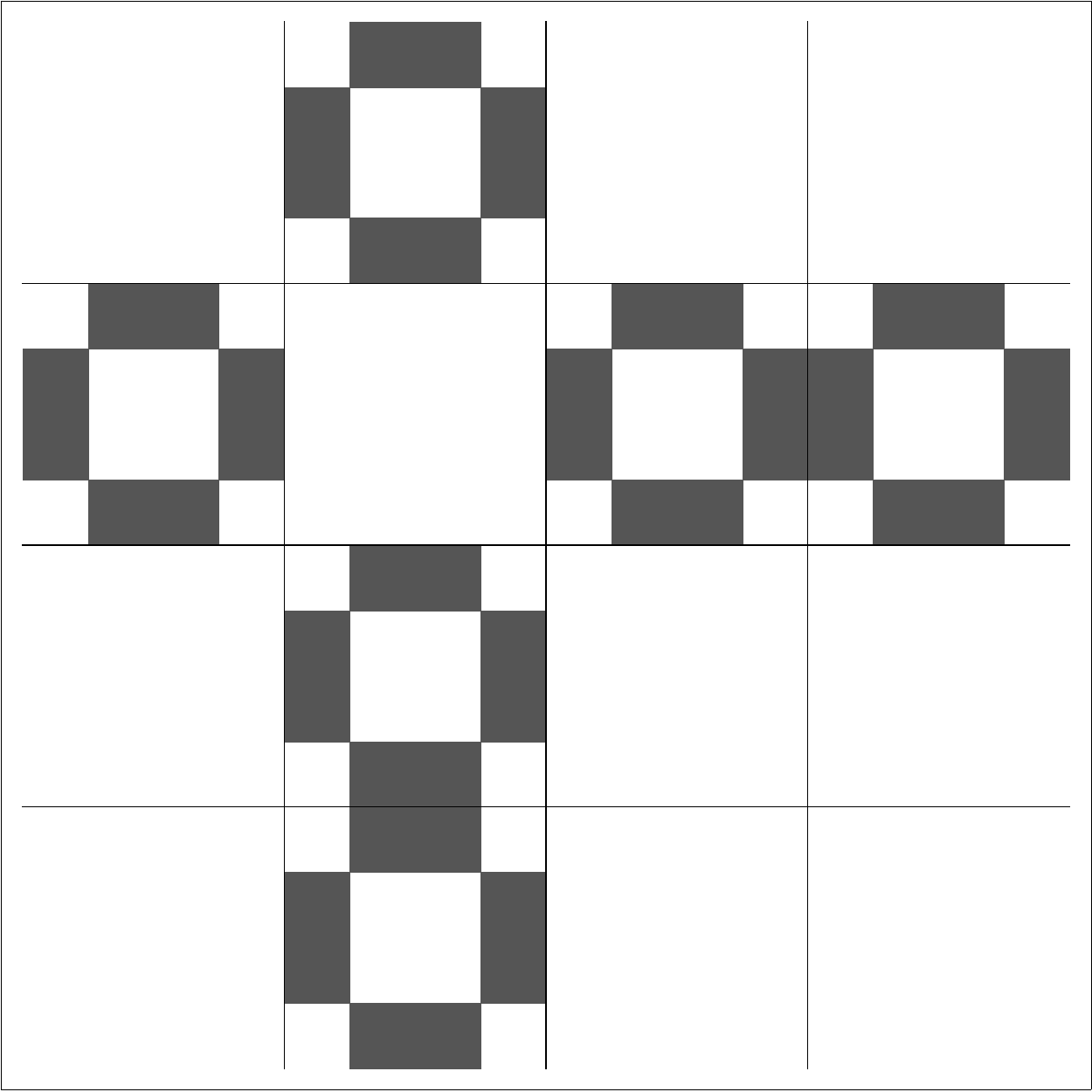}}\right)
\end{array}$
\end{minipage}\\
&\\
(c) Strong product $G \SP H$ & \\
\hspace*{0.6in}\raisebox{-0.5\height}{\includegraphics[scale=0.39]{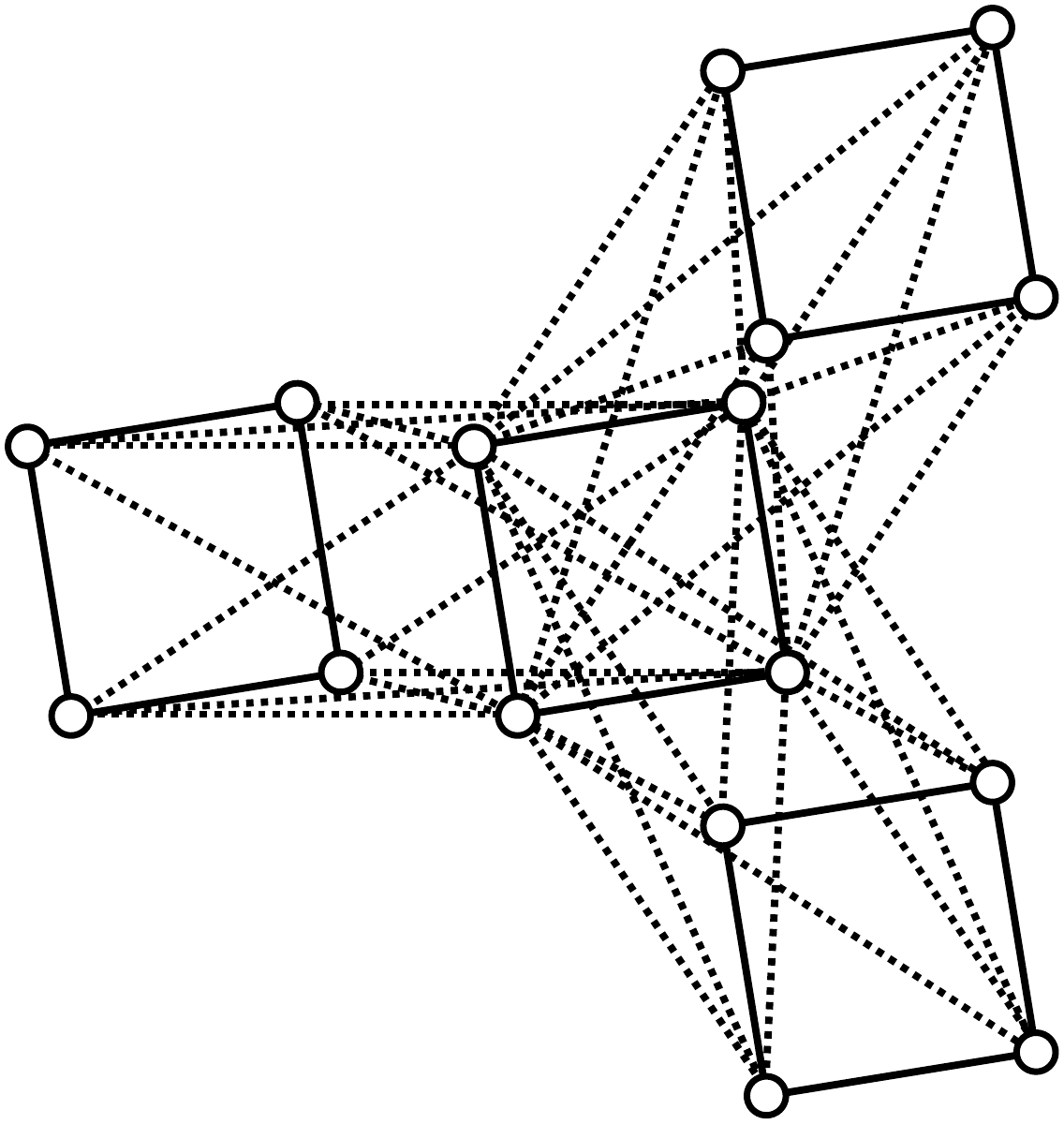}} &
\begin{minipage}{3in}
$\begin{array}{rl}%
A_{G \SP H} &= A_G \KS A_H + A_G \KP A_H\\
&=\left(\raisebox{-0.48\height}{\includegraphics[scale=0.29]{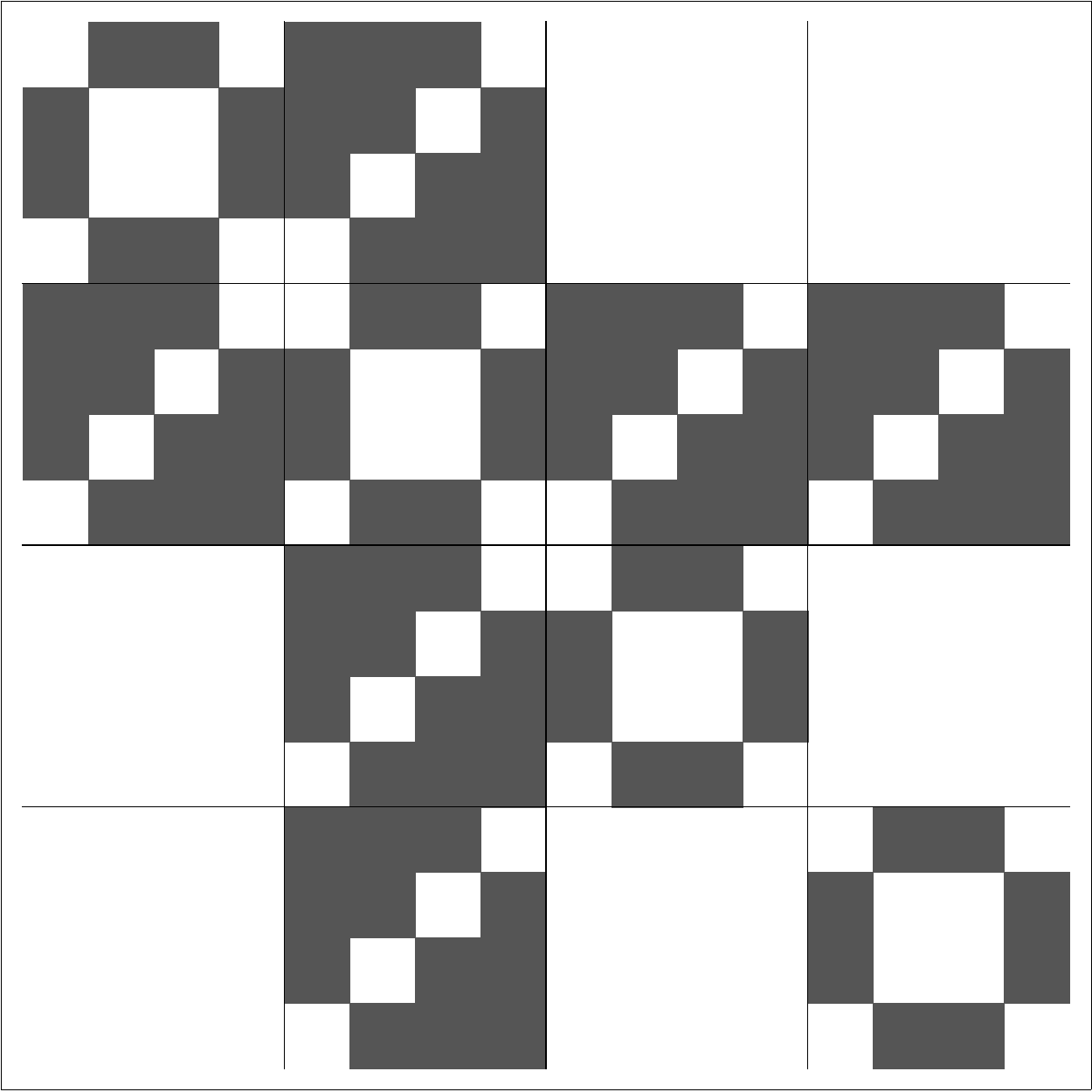}}\right)
\end{array}$
\end{minipage}
\end{tabular}
\caption{Examples of three graph products. Top: Two factor graphs used
  in this example, $G$ and $H$, and their adjacency matrices. (a)
  Cartesian product. (b) Direct (tensor) product. (c) Strong
  product. In graphically presented adjacency matrices, gray and white
  blocks represent 1s and 0s, respectively. Here, $G$ is considered an
  inter-layer network while $H$ is considered an intra-layer network,
  but the opposite interpretation is also possible.}
\label{fig:GPMN-examples}
\end{figure}

Three graph product operations have been considered so far: Cartesian
product, direct (tensor) product, and strong product
\cite{sayama2016estimation}. We will also discuss their generalization
later in this paper. Let $G=(V_G, E_G)$ and $H=(V_H, E_H)$ be two
simple graphs as factor networks, where $V_G$ (or $V_H$) and $E_G$ (or
$E_H$) are the sets of nodes and edges of $G$ (or $H$), respectively.
Also let $A_G$ and $A_H$ be the adjacency matrices of $G$ and $H$,
respectively. For all of the aforementioned three graph products, a
new node set is given by a Cartesian product of $V_G$ and $V_H$, i.e.,
$\{(g,h) \; | \; g\in V_G , \, h\in V_H\}$. Then the three graph
products are defined as follows:
\begin{description}
\item[Cartesian product $G \CP H$] A network in which two nodes $(g,
  h)$ and $(g', h')$ are connected if and only if $g = g'$ and $(h,
  h') \in E_H$, or $h = h'$ and $(g, g') \in E_G$. Its adjacency
  matrix is given by
\begin{equation}
A_{G \CP H} = A_G \KS A_H = A_G \KP I_{|V_H|} + I_{|V_G|} \KP A_H, \label{eq:cp}
\end{equation}
where $\KS$ and $\KP$ are Kronecker sum and Kronecker product
operators, respectively, and $I_n$ is an $n \times n$ identity matrix.
\item[Direct (tensor) product $G \DP H$] A network in which two nodes
  $(g, h)$ and $(g', h')$ are connected if and only if $(g, g') \in
  E_G$ and $(h, h') \in E_H$. Its adjacency matrix is given by
\begin{equation}
A_{G \DP H} = A_G \KP A_H. \label{eq:dp}
\end{equation}
\item[Strong product $G \SP H$] A network that is obtained as the sum
  of $G \CP H$ and $G \DP H$. Its adjacency matrix is given by
\begin{equation}
A_{G \SP H} = A_G \KS A_H + A_G \KP A_H. \label{eq:sp}
\end{equation}
\end{description}
Examples are shown in Figure \ref{fig:GPMN-examples}. We call these
three types of GPMNs {\em Cartesian product multilayer networks
  (CPMNs)}, {\em direct product multilayer networks (DPMNs)}, and {\em
  strong product multilayer networks (SPMNs)}. The number of nodes is
$|V_G| |V_H|$ for all of these three graph products. CPMNs guarantee
that all inter-layer edges are diagonal and layer-coupled (i.e.,
independent of the intra-layer nodes), and therefore, all CPMNs are
also multiplex networks (in the sense that all layers share the same
intra-layer node set and the inter-layer edges are all diagonal)
\cite{kivela2014multilayer}. Some other known topological properties
are summarized in Table \ref{tab:known-properties}.

\begin{table}[tp]
\centering
\caption{Some of known topological properties of GPMNs with simple,
  undirected factor networks $G$ and $H$. $\langle k_G \rangle$ and
  $\langle k_H \rangle$ represent the average degrees of $G$ and $H$,
  respectively.}
\begin{tabular}{c|cccc}
\hline
Type & \# of nodes & \# of edges & Average degree & Multiplex? \\
\hline
&&&&\\
CPMN $G \CP H$ & $|V_G| |V_H|$ & $|E_G||V_H| + |V_G||E_H|$ &
$\displaystyle
\frac{2|E_G|}{|V_G|} + \frac{2|E_H|}{|V_H|} = \langle k_G \rangle + \langle k_H \rangle 
$ & Yes \\
&&&&\\
DPMN $G \DP H$ & $|V_G| |V_H|$ & $2|E_G||E_H|$ &
$\displaystyle
\frac{4|E_G||E_H|}{|V_G| |V_H|} = \langle k_G \rangle \langle k_H \rangle 
$ & No \\
&&&&\\
SPMN $G \SP H$ & $|V_G| |V_H|$ & $|E_G||V_H| + |V_G||E_H| + 2|E_G||E_H|$ &
$\displaystyle\begin{array}{l}
\frac{2|E_G|}{|V_G|} + \frac{2|E_H|}{|V_H|} + \frac{4|E_G||E_H|}{|V_G| |V_H|} \\
= \langle k_G \rangle + \langle k_H \rangle + \langle k_G \rangle \langle k_H \rangle 
\end{array}$ & No \\
&&&&\\
\hline
\end{tabular}
\label{tab:known-properties}
\end{table}

The three graph product operators described above are commutative,
i.e., the resulting graphs of $G * H$ and $H * G$ (where $*$ is either
$\CP$, $\DP$, or $\SP$) are isomorphic to each other with appropriate
node permutations. Therefore, it is an arbitrary choice which factor
network is considered intra- or inter-layer. In addition, these graph
product operators are also known to be associative.

One interesting, and quite useful, fact already known about GPMNs is
that the spectral properties of their degree, adjacency, and Laplacian
matrices are related to those of their factor networks
\cite{macduffee1933theory,fiedler1973algebraic,sayama2016estimation}. More
specifically, degree and adjacency spectra of CP/DP/SPMNs and
Laplacian spectra of CPMNs are characterized exactly by the spectra of
their factor networks, as follows \cite{sayama2016estimation}:
\begin{description}
\item[Degree spectra] With $\left(d^G_i\right)$ and
  $\left(d^H_j\right)$ being the degree spectra (node degrees) of factor networks $G$
  and $H$, respectively:
\begin{description}
\item[CPMN] $\left(d^G_i + d^H_j\right)$ ~~$\forall i, j$
\item[DPMN] $\left(d^G_i d^H_j\right)$ ~~$\forall i, j$
\item[SPMN] $\left(d^G_i + d^H_j + d^G_i d^H_j\right)$ ~~$\forall i, j$ 
\end{description}
\item[Adjacency spectra] With $\left(\lambda^G_i\right)$ and
  $\left(\lambda^H_j\right)$ being the adjacency spectra (eigenvalues
  of adjacency matrices) of factor networks $G$ and $H$, respectively:
\begin{description}
\item[CPMN] $\left(\lambda^G_i + \lambda^H_j\right)$ ~~$\forall i, j$
\item[DPMN] $\left(\lambda^G_i \lambda^H_j\right)$ ~~$\forall i, j$
\item[SPMN] $\left(\lambda^G_i + \lambda^H_j + \lambda^G_i \lambda^H_j\right)$ ~~$\forall i, j$ 
\end{description}
\item[Laplacian spectra] With $\left(\mu^G_i\right)$ and
  $\left(\mu^H_j\right)$ being the Laplacian spectra (eigenvalues
  of Laplacian matrices) of factor networks $G$ and $H$, respectively:
\begin{description}
\item[CPMN] $\left(\mu^G_i + \mu^H_j\right)$ ~~$\forall i, j$
\end{description}
\end{description}

In contrast to these, no exact formula is known for the
characterization of Laplacian spectra of DP/SPMNs using their factor
networks' Laplacian spectra. This is because the Laplacian matrices of
DP/SPMNs cannot be expressed simply by using the Laplacian matrices of
their factor networks, but instead, they involve the Kronecker
products of degree and Laplacian matrices of the factor networks
\cite{sayama2016estimation}. Such a coupling between degree and
Laplacian matrices makes it hard to obtain Laplacian spectra of
DP/SPMNs analytically.

However, we have recently reported \cite{sayama2016estimation} that
the Laplacian spectra of DP/SPMNs can still be approximated
heuristically using degree and Laplacian spectra of their factor
networks, as follows:
\begin{description}
\item[Approximated Laplacian spectra] With $\left[\left(d^G_i\right),
  \left(\mu^G_i\right)\right]$ and $\left[\left(d^H_j\right),
  \left(\mu^H_j\right)\right]$ being the degree/Laplacian spectra of
  factor networks $G$ and $H$, respectively:
\begin{description}
\item[DPMN] $\left(\mu^G_i d^H_j + d^G_i \mu^H_j - \mu^G_i \mu^H_j \right)$ ~~$\forall i, j$
\item[SPMN] $\left(\mu^G_i + \mu^H_j + \mu^G_i d^H_j + d^G_i \mu^H_j - \mu^G_i \mu^H_j \right)$ ~~$\forall i, j$
\end{description}
\end{description}
Note that these formulae involve two distinct spectra for each factor
network, $\left( d \right)$ and $\left( \mu \right)$, whose orderings
are independent from each other. This implies that optimizing their
orderings can help improve the accuracy of the heuristic
approximation. In \cite{sayama2016estimation}, we explored several
different ordering methods and found that sorting both spectra in
an ascending order achieves the best approximation (we will revisit
this issue in the next section). See \cite{sayama2016estimation} for
more details.

In summary, the spectral relationships between GPMNs and their factor
networks described above help study structural and dynamical
properties of GPMNs, such as degree distributions, bipartiteness,
algebraic connectivities, number of connected components, spectral
gaps, eigenratios, and so on.

\section{Spectral Properties of Nonsimple GPMNs}
\label{sec:nonsimple}

In this paper, we aim to extend the definitions of GPMNs to make them
capable of capturing a greater variety of complex networks.

The first extension is to consider nonsimple graphs with directed,
weighted, signed, and/or self-looped edges. Mathematically, this
relaxation is equivalent to considering any arbitrary real-valued
square matrices for adjacency matrices of factor networks, $A_G$ and
$A_H$. Through the rest of the paper, ``adjacency matrices'' are used
in this broader definition. Note that the ways adjacency matrices of
graph products are calculated in Eqs.~(\ref{eq:cp}), (\ref{eq:dp}),
and (\ref{eq:sp}) do not assume graph simplicity, so they can
seamlessly apply to nonsimple graphs as is. We call the resulting
networks {\em nonsimple GPMNs}.

Of particular interest is whether the spectral relationships between
simple GPMNs and their factor networks described in Section
\ref{sec:simple} also apply to nonsimple GPMNs. For asymmetric
networks with weighted/signed edges, we define strength (degree) and
Laplacian matrices as follows (note that these Laplacian matrices are
no longer symmetric or positive-semidefinite in general):
\begin{description}
\item[In-strength matrix] $D^\mathrm{in} = \mathrm{diag}(A \bone_n)$
\item[Out-strength matrix] $D^\mathrm{out} = \mathrm{diag}(\bone^T_n A)$
\item[In-strength Laplacian matrix] $L^\mathrm{in} = D^\mathrm{in} - A$
\item[Out-strength Laplacian matrix] $L^\mathrm{out} = D^\mathrm{out} - A$
\end{description}
Here $A$ is an $n \times n$ asymmetric, weighted and/or signed
adjacency matrix with possible self-loops (non-zero diagonal entries),
and $\bone_n$ is an all-one column vector of size $n$.

We take note that the analytical characterizations of degree and
adjacency spectra of simple CP/DP/SPMNs and Laplacian spectra of CPMNs
\cite{sayama2016estimation} were solely based on the algebraic
properties of Kronecker sum and product, and therefore they also apply
to nonsimple counterparts without any modification (we also confirmed
this numerically; results not shown).

Some additional considerations are needed for Laplacian spectra of
nonsimple DP/SPMNs. As described in Section \ref{sec:simple}, the
original approximation method developed for simple DP/SPMNs
\cite{sayama2016estimation} involved the sorting of degree and
Laplacian spectra of factor networks. This requires additional
investigation when applied to nonsimple networks, because adjacency
and Laplacian matrices of nonsimple networks are generally asymmetric
and thus have complex eigenvalues, for which sorting is not obvious.

\begin{figure}[t]
\centering
\begin{tabular}{lcl}
~~(a) DPMNs & ~~~~ & ~~(b) SPMNs \\
\includegraphics[width=0.44\textwidth]{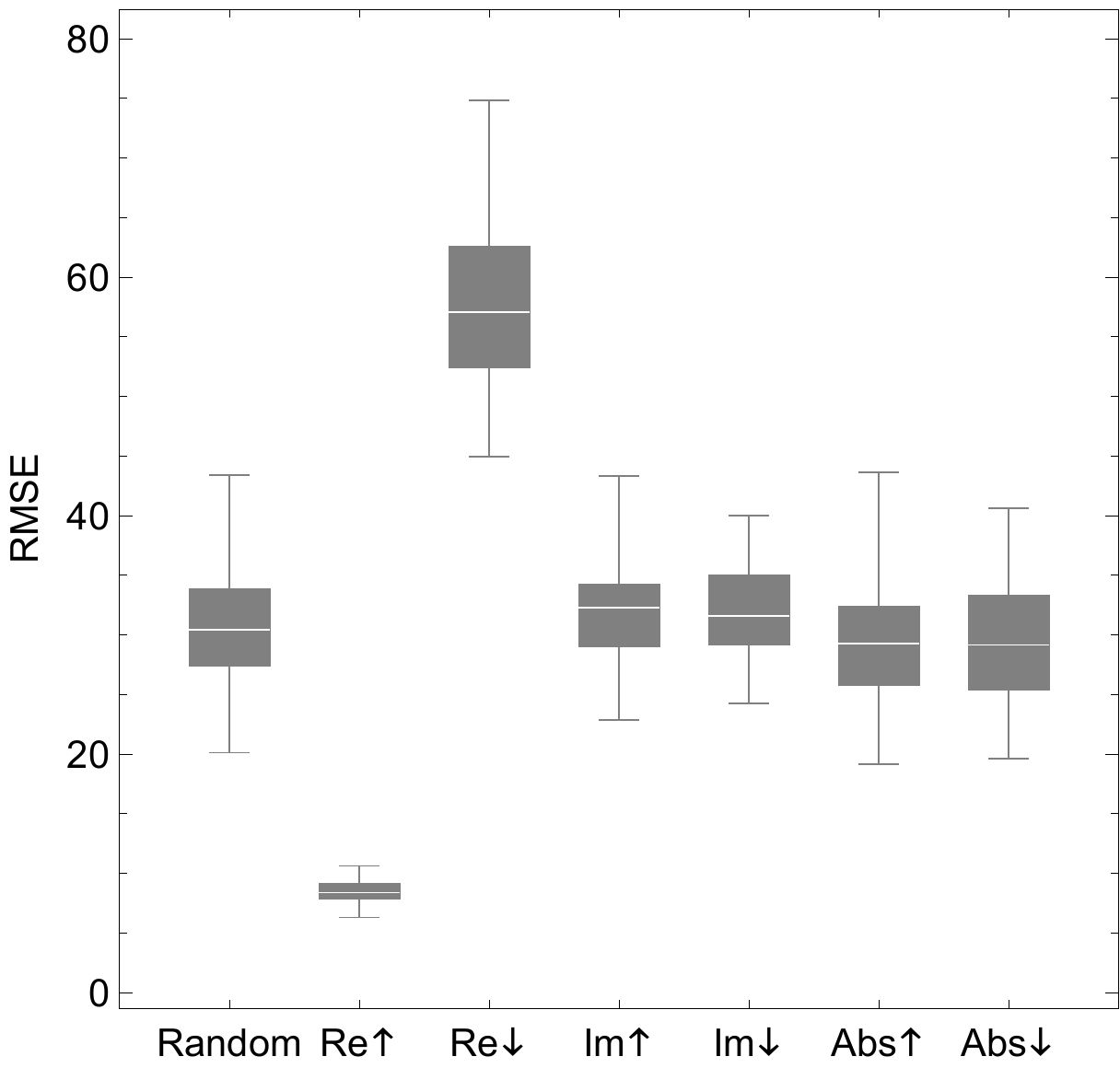} & ~~~~ &
\includegraphics[width=0.44\textwidth]{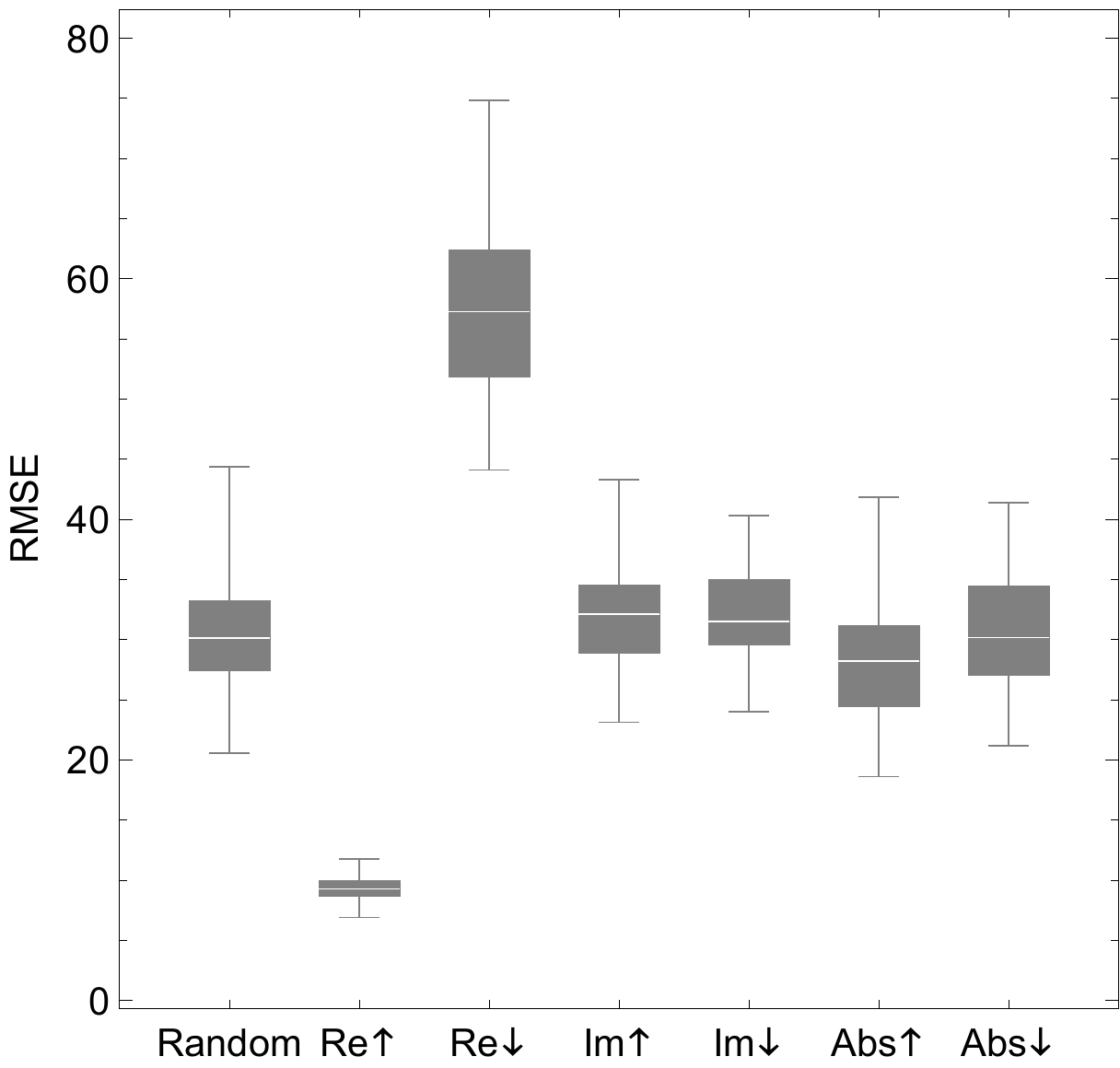} \\
\end{tabular}
\caption{Comparison of approximation performance among seven heuristic
  eigenvalue-sorting criteria (see text for details) for approximating
  in-strength Laplacian spectra of nonsimple (a) DPMNs and (b)
  SPMNs. ``$\uparrow$'' and ``$\downarrow$'' denote ascending and
  descending orders, respectively. The root mean square error (RMSE)
  per eigenvalue was used as a performance metric, in which an error
  was measured by the absolute value of the difference between true
  and approximated eigenvalues. Results were collected from one
  hundred independent tests for each condition, and their
  distributions were shown in box-whisker plots. Two adjacency
  matrices of factor networks used were: $A_G=$ a random matrix whose
  entries were randomly sampled from a uniform distribution $[-1,1]$,
  and $A_H=$ a random matrix whose entries were randomly sampled from
  a uniform distribution $[-2,2]$. The sizes of $A_G$ and $A_H$ were
  randomly and independently set between 40 and 60. These parameter
  values were chosen arbitrarily just for illustrative purposes, while
  the overall trends of results were robust to parameter
  variations. ANOVA and Tukey/Bonferroni posthoc tests showed
  extremely significant differences among the conditions ($p=3.75
  \times 10^{-331}$ for DPMNs; $p=2.13 \times 10^{-324}$ for
  SPMNs). Results for out-strength Laplacian spectra showed similar
  trends.}
\label{fig:laplacian-approximation-nonsimple-sortings}
\end{figure}

\begin{figure}[t]
\centering
\begin{tabular}{lcl}
~~(a) DPMN & ~~~~ & ~~(b) SPMN \\
\includegraphics[width=0.44\textwidth]{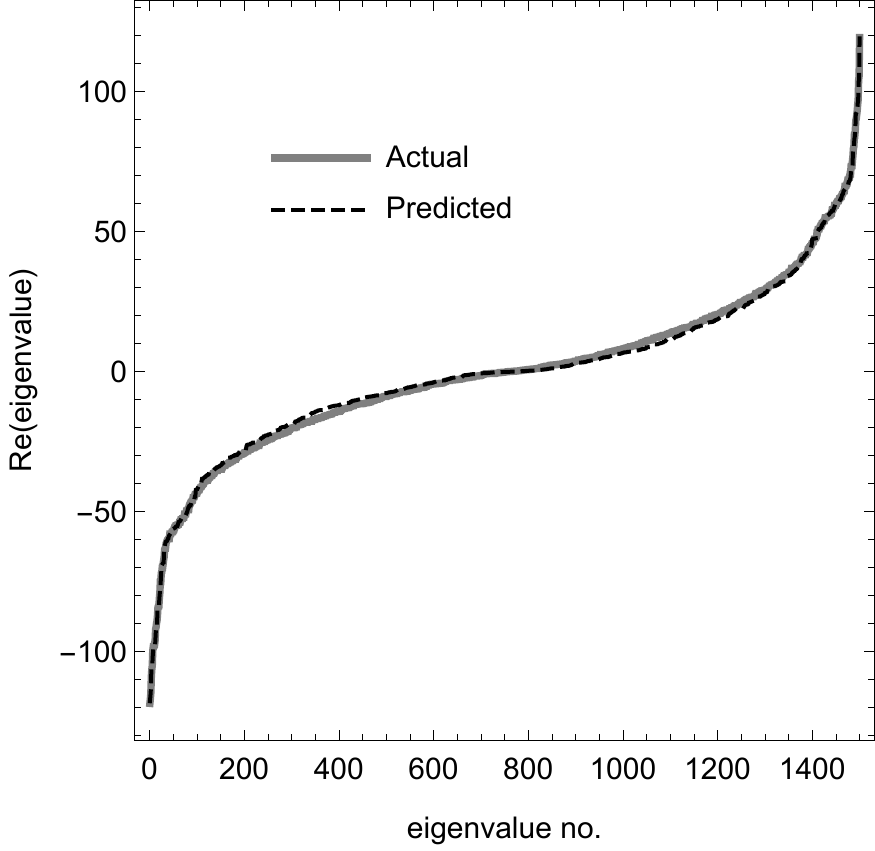} & ~~~~ &
\includegraphics[width=0.44\textwidth]{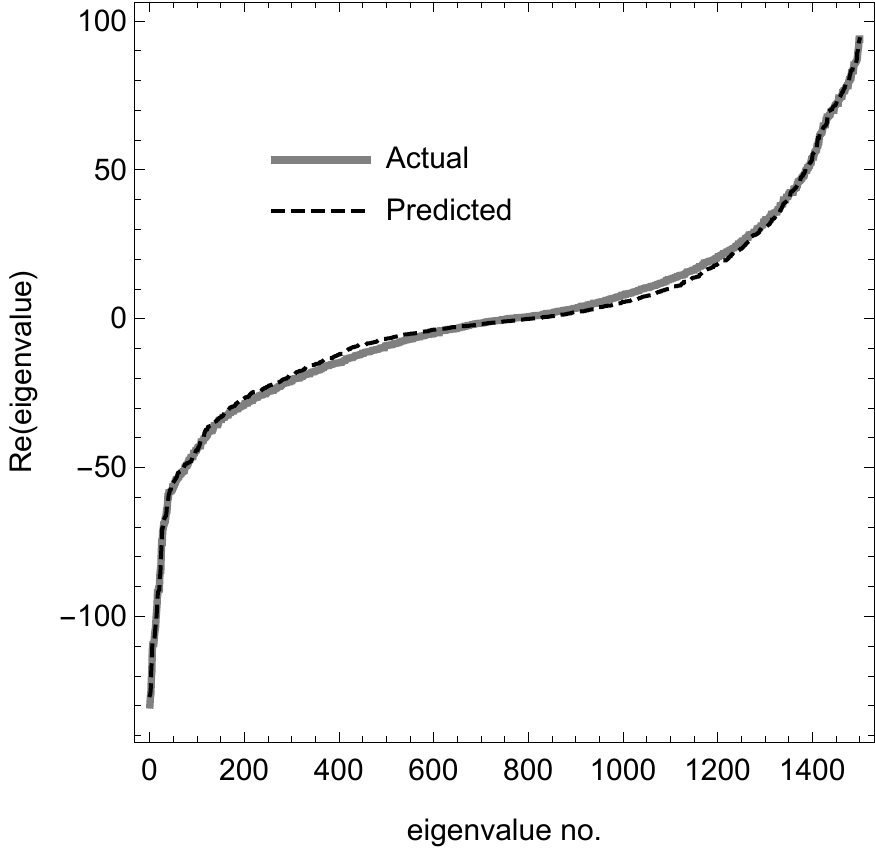} \\
\end{tabular}
\caption{Examples of in-strength Laplacian spectra of (a) a nonsimple
  DPMN and (b) a nonsimple SPMN, each approximated using the same
  methods as described in Section \ref{sec:simple}
  \cite{sayama2016estimation} (black dashed curves), in comparison
  with the actual ones (gray solid curves). Eigenvalues are sorted by
  their real parts in an ascending order, and only their real parts
  are plotted (while imaginary parts were more or less randomly
  distributed). Two adjacency matrices of factor networks used were:
  $A_G=$ a $50 \times 50$ random matrix whose entries were randomly
  sampled from a uniform distribution $[-1,1]$, and $A_H=$ a $30
  \times 30$ random matrix whose entries were randomly sampled from a
  uniform distribution $[-2,2]$. These parameter values were chosen
  arbitrarily just for illustrative purposes, while the overall trends
  of results were robust to parameter variations. Following the
  results shown in
  Fig.~\ref{fig:laplacian-approximation-nonsimple-sortings}, the
  factor networks' node strengths and Laplacian eigenvalues were
  sorted in ascending order (by their real parts for the latter)
  before the approximation method was used.}
\label{fig:laplacian-approximation-nonsimple}
\end{figure}

To address this issue, we have conducted numerical experiments with
the following seven different heuristic criteria for sorting Laplacian
eigenvalues of factor networks (while node strengths were always
sorted in an ascending order):
\begin{enumerate}
\item Random
\item By real part (ascending)
\item By real part (descending)
\item By imaginary part (ascending)
\item By imaginary part (descending)
\item By absolute value (ascending)
\item By absolute value (descending)
\end{enumerate}
Results are summarized in
Fig.~\ref{fig:laplacian-approximation-nonsimple-sortings}. The results
clearly show that, when the eigenvalues of Laplacian matrices of
factor networks are sorted by their real parts in an ascending order
(which is the most natural generalization of the sorting method used
in \cite{sayama2016estimation}), the same methods described in Section
\ref{sec:simple} can approximate in- and out-strength Laplacian
spectra of nonsimple DP/SPMNs most effectively. Examples of
approximations are shown in
Fig.~\ref{fig:laplacian-approximation-nonsimple}.

To summarize, the extension of GPMNs to nonsimple networks can be done
in a straightforward manner, and the known spectral relationships
between GPMNs and their factor networks are also applicable to
nonsimple ones with little to no modification.

\section{Generalizing Graph Product}
\label{sec:generalized-gp}

The second extension being made in this paper is to generalize graph
product operation to arbitrary linear combination of Cartesian and
direct products. Specifically, we define the following {\em
  generalized product}:
\begin{description}
\item[Generalized product $\displaystyle G \GP_{\alpha, \beta} H$] A
  network that is obtained as a weighted linear combination of $G \CP
  H$ and $G \DP H$, where $\alpha$ and $\beta$ are used as
  weights. Its adjacency matrix is given by
\begin{equation}
A_{G \GP_{\alpha, \beta} H} = A_G \KG_{\alpha, \beta} A_H = \alpha A_G \KS A_H + \beta A_G \KP A_H, \label{eq:gp}
\end{equation}
\end{description}
where the symbol $\displaystyle\KG_{\alpha, \beta}$ is newly
introduced in this paper to represent a weighted sum of the Kronecker
sum and product. This graph product operation is still commutative and
associative. All of CP/DP/SPMNs can be described uniformly using this
generalized product ($\displaystyle\GP_{1,0} = \CP$,
$\displaystyle\GP_{0,1} = \DP$, and $\displaystyle\GP_{1,1} =
\SP$). Moreover, setting $\alpha$ and/or $\beta$ to non-integer values
represents a more nuanced balance between Cartesian and direct
products in the GPMN structure (note that this is made possible by the
first extension). We call networks generated by using this generalized
product {\em generalized GPMNs (GGPMNs)}.

We have found that the strength (degree), adjacency, and Laplacian
spectra of GGPMNs can be given exactly or approximately as follows:
\begin{description}
\item[In- and out-strength spectra (= in- and out-strength sequence)]
  With $\left( d^G_i \right)$ and $\left( d^H_j \right)$ being the in-
  or out-strength spectra (node strengths) of factor networks $G$ and
  $H$, respectively:
\begin{description}
\item[GGPMN] $\left( \alpha d^G_i + \alpha d^H_j + \beta d^G_i d^H_j \right)$ ~~ $\forall i, j$
\end{description}
\item[Adjacency spectra] With $\left( \lambda^G_i \right)$ and $\left(
  \lambda^H_j \right)$ being the adjacency spectra (eigenvalues of
  adjacency matrices) of factor networks $G$ and $H$, respectively:
\begin{description}
\item[GGPMN] $\left( \alpha \lambda^G_i + \alpha \lambda^H_j + \beta \lambda^G_i \lambda^H_j \right)$
~~ $\forall i, j$
\end{description}
\item[Approximated Laplacian spectra] With $\left[\left(d^G_i\right),
  \left(\mu^G_i\right)\right]$ and $\left[\left(d^H_j\right),
  \left(\mu^H_j\right)\right]$ being the in- or out-strength/Laplacian
  spectra of factor networks $G$ and $H$, respectively (all spectra
  should be sorted in an ascending order of real parts for best
  approximation):
\begin{description}
\item[GGPMN] $\left( \alpha \mu^G_i + \alpha \mu^H_j + \beta \mu^G_i d^H_j + \beta d^G_i \mu^H_j - \beta \mu^G_i \mu^H_j \right)$ ~~ $\forall i, j$
\end{description}
\end{description}
Details of derivation are given in Appendix. Note that these spectral
relationships described in this section hold for both simple and
nonsimple GGPMNs.

With these two extensions introduced in Sections \ref{sec:nonsimple}
and \ref{sec:generalized-gp}, GPMNs as a modeling framework have
gained an enhanced expressive power and can be used to describe a
wider variety of complex networks. In the following sections, we
illustrate the use of GPMNs through three application examples.

\section{Application I: Predicting Epidemic Thresholds}
\label{sec:app1}

GPMNs can be used to predict epidemic thresholds (i.e., critical ratio
of infection and recovery rates) of epidemic processes on networks. It
is known that the epidemic threshold is given by the reciprocal of the
largest eigenvalue of the adjacency matrix of the network
\cite{chakrabarti2008epidemic}. If the network can be modeled as a
GGPMN, the largest eigenvalue of its adjacency matrix can be obtained
analytically as
\begin{equation}
\lambda_{\max}^{G \GP_{\alpha, \beta} H} = \max_{i,j} \left( \alpha \lambda^G_i + \alpha \lambda^H_j + \beta \lambda^G_i \lambda^H_j \right) . \label{eq:max-eigval}
\end{equation}
In most cases, the adjacency spectrum of a large simple graph is
characterized by a dense area around the origin accompanied by a small
number of substantially larger (positive) outliers
\cite{farkas2001spectra}. Therefore, if $\alpha \ge 0$ and $\beta \ge
0$, Eq.~(\ref{eq:max-eigval}) is simplified to
\begin{equation}
\lambda_{\max}^{G \GP_{\alpha, \beta} H} = \alpha \lambda^G_{\max} + \alpha \lambda^H_{\max} + \beta \lambda^G_{\max} \lambda^H_{\max} . \label{eq:lambdamax-formula}
\end{equation}
Because the epidemic threshold is the reciprocal of the largest
eigenvalue, this result indicates that the epidemic threshold on a
CPMN ($(\alpha, \beta) = (1, 0)$) is twice the harmonic mean of the
thresholds on two factor networks. Similarly, the epidemic threshold
on a DPMN ($(\alpha, \beta) = (0, 1)$) is simply the product of the
thresholds on two factor networks.

\begin{figure}
\centering
\includegraphics[width=.55\columnwidth]{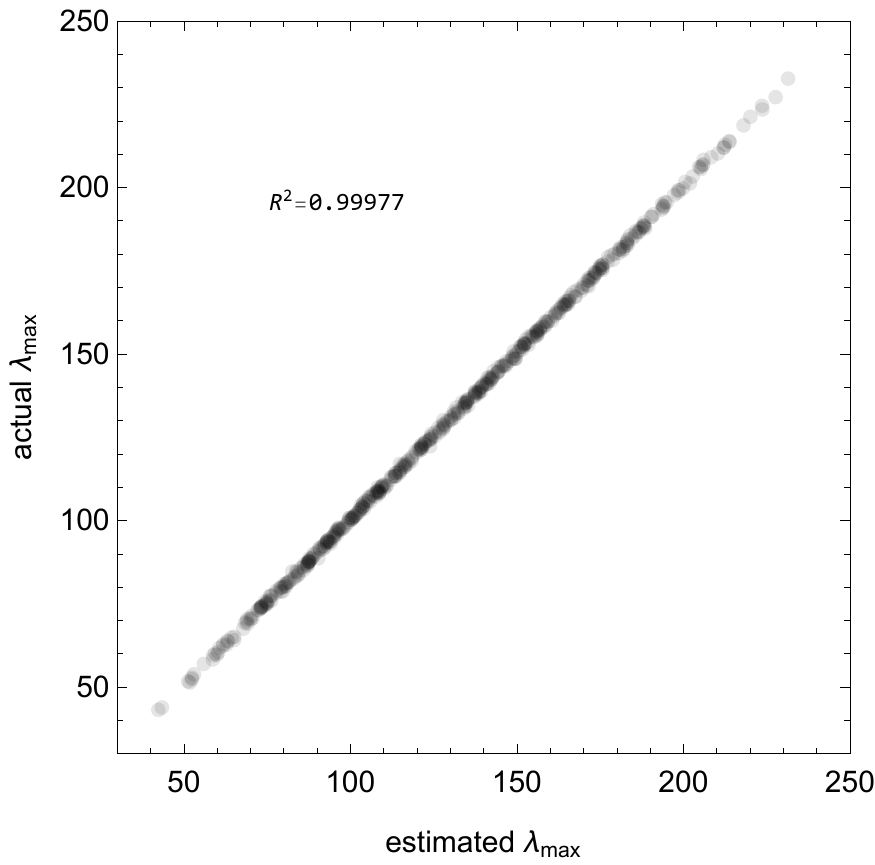}
\caption{Comparison of estimated and actual largest eigenvalues of
  adjacency matrices of random networks generated by a stochastic
  block model. Results were collected from five hundred independent
  calculations. In each case, the number of communities ($r$) and the
  size of each community ($m$) were randomly chosen between 3 and 7
  and between 40 and 60, respectively.  These parameter values were
  chosen arbitrarily just for illustrative purposes, while the overall
  trends of results were robust to parameter variations. $P$ and an
  actual binary adjacency matrix were randomly generated to be
  symmetric in each case. The estimated values were calculated using
  Eq.~(\ref{eq:threshold-estimation}), while the actual ones were
  explicitly computed from the binary adjacency matrices.}
\label{fig:threshold-estimation}
\end{figure}

A particularly interesting application of this result is the
prediction of epidemic thresholds on random networks generated by a
stochastic block model with equal community size. Let $P$ be an $r
\times r$ edge probability matrix, and assume each community is made
of exactly $m$ nodes (thus the total number of nodes is $rm$). Then,
the ensemble average of adjacency matrices of networks generated by
this model is given by $P \KP R$, an adjacency matrix of a nonsimple
DPMN, where $R$ is an $m \times m$ all-one square matrix. A typical
assumption in matrix perturbation theory suggests that the spectrum of
a specific binary adjacency matrix generated from this stochastic
block model behaves similarly with the spectrum of this ensemble
average. Therefore, we can estimate the epidemic threshold on a random
network generated with this model by calculating the largest
eigenvalue of $P \KP R$. Since $R$ has $m$ as the single largest
eigenvalue (while all other eigenvalues are 0s), this estimation
becomes as simple as
\begin{align}
\lambda_{\max}^{P \KP R} = \lambda^P_{\max} m . \label{eq:threshold-estimation}
\end{align}
This accomplishes a significant reduction of computational cost
because the number of communities (size of $P$) is usually
significantly smaller than the size of the entire network. Figure
\ref{fig:threshold-estimation} presents results of a numerical
experiment that compared the estimated largest eigenvalues obtained
using this method with the actual ones, showing excellent matching
between the two.

We note that other recent studies also discuss spectral properties of
stochastic block models, mainly from the viewpoint of community
detectability
\cite{nadakuditi2012graph,peixoto2013eigenvalue,ghasemian2016detectability}. Compared
to them, the example presented in this section above is unique in
using graph product as a concise mathematical representation of
network topology and focusing specifically on the prediction of
epidemic thresholds.

\section{Application II: Propagation in Nontrivial Space and Time}
\label{sec:app2}

GPMNs can be used to represent dynamical processes taking place in
nontrivial spatial and/or temporal structures. A popular way of doing
this is to use one of the factor networks as a network-shaped spatial
structure, while the other as an identical dynamical network that is
embedded inside of each node in space. There are several such
multilayer network models developed and used in the literature,
including coupled cellular networks \cite{golubitsky2009bifurcations}
and ecological networks on interconnected habitats
\cite{brechtel2016master}. In these cases, network topologies that can
be produced by Cartesian products were typically used to couple
multiple identical intra-layer networks over space.

Here, we illustrate other uses of GPMNs to represent dynamical
processes that involve not just nontrivial space but also
time. Specifically, we use graph product to represent various types of
weighted random walks on a network. Let $G$ be a weighted, directed
network that represents transition likelihoods, and $T$ a chain made
of directed edges (Fig.~\ref{fig:GPMN-random-walk}, top). Each of
these networks in isolation can be considered a representation of
either spatial movement without time ($G$), or the flow of time
without spatial movement ($T$). Applying different graph product
operators to these two factor networks produces a specific
spatio-temporal representation of different random walk processes, as
follows:

\begin{figure}
\centering
\begin{tabular}{ll}
$G$ : & \raisebox{-0.5\height}{\includegraphics[scale=0.5]{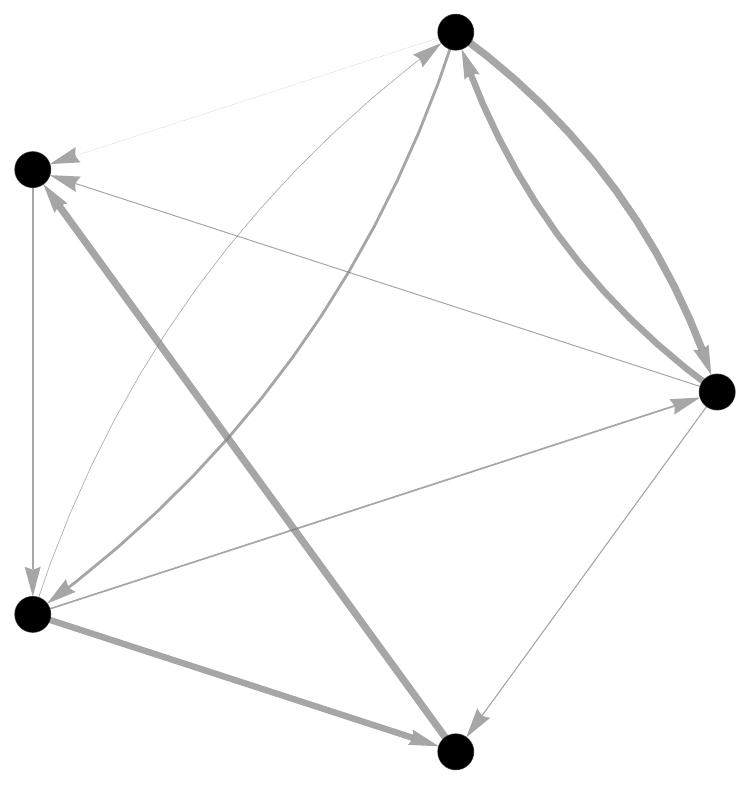}} \\
~ & ~\\
$T$ : & \raisebox{-0.3\height}{\includegraphics[scale=0.5]{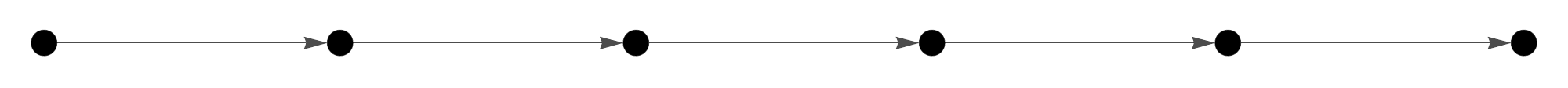}} \\
~ & ~\\
\end{tabular}\\
\begin{tabular}{l}
(a) $G \DP T$ :\\
\includegraphics[width=0.9\textwidth]{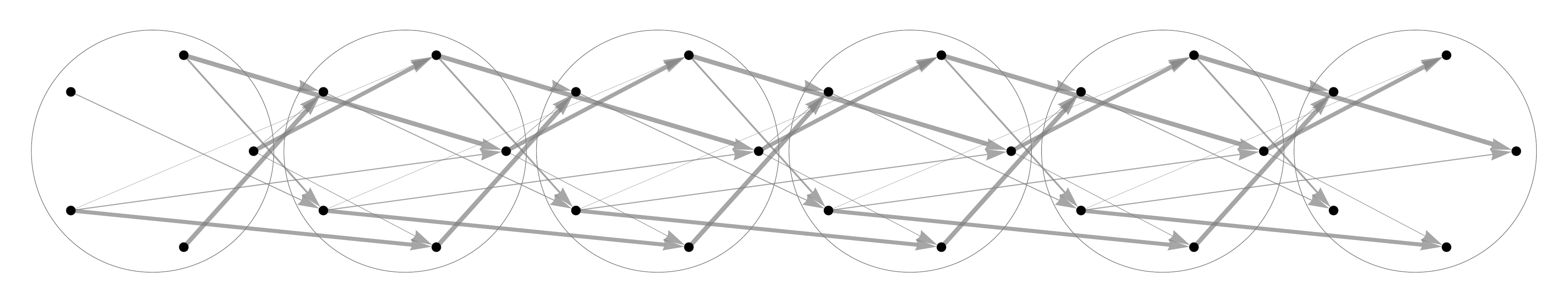}\\
~\\
(b) $G \CP T$ :\\
\includegraphics[width=0.9\textwidth]{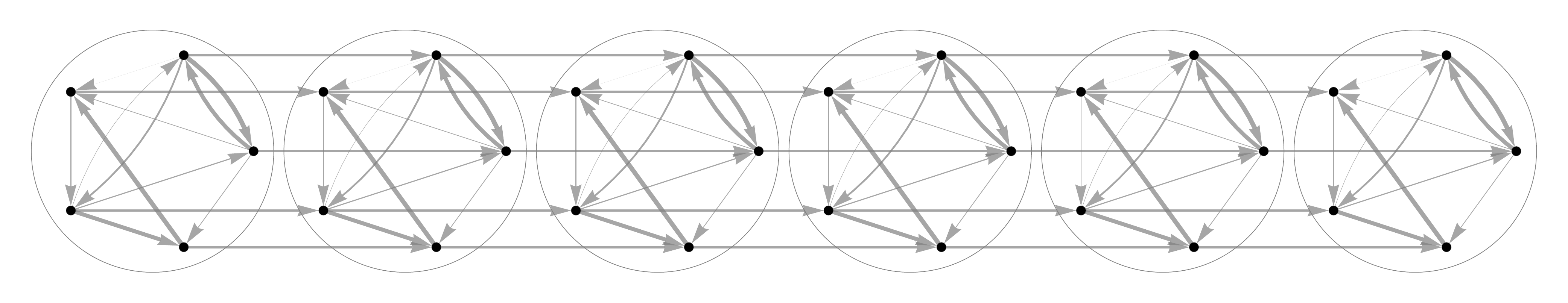}\\
~\\
(c) $\displaystyle G \GP_{\alpha, \beta} T$ :\\
\includegraphics[width=0.9\textwidth]{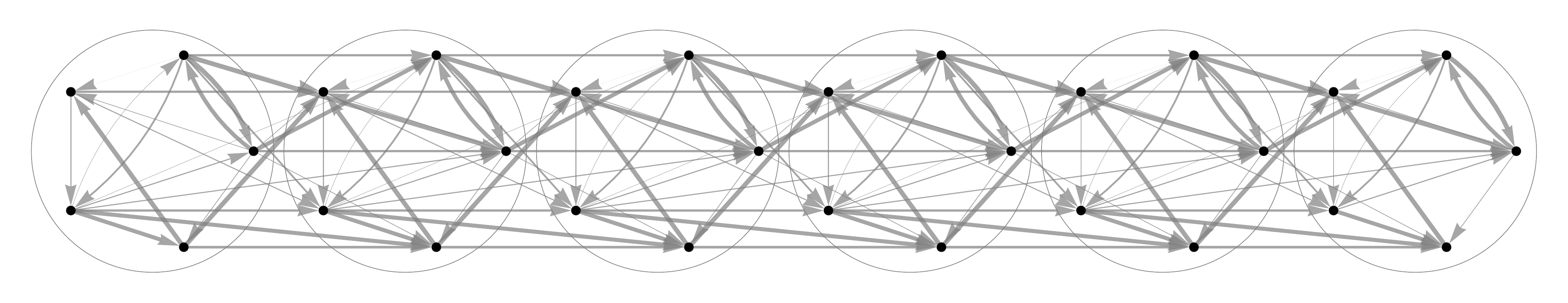}\\
\end{tabular}
\caption{Weighted random walk processes represented by GPMNs. Top: Two
  factor networks. $G$ represents an underlying spatial transition
  network, while $T$ represents time. (a) Direct product $G \DP T$,
  representing random walk with forced temporal progress. (b)
  Cartesian product $G \CP T$, representing random walk within each
  time segment with occasional stochastic temporal progress. (c)
  Generalized product $\displaystyle G \GP_{\alpha, \beta} T$,
  representing the combination of (a) and (b). Circles represent
  layers (time segments).}
\label{fig:GPMN-random-walk}
\end{figure}

\begin{description}
\item[Direct product $G \DP T$] This represents weighted random walk
  with forced temporal progress
  (Fig.~\ref{fig:GPMN-random-walk}(a)). Each layer contains no
  intra-layer edges, and all the edges connect a node in one layer to
  another node in the subsequent layer, indicating that no entities
  can stay within the same time point when they make transitions. This
  is the most natural representation of ``vanilla'' random walk in
  which time is explicitly represented by $T$.
\item[Cartesian product $G \CP T$] This represents weighted random
  walk taking place within each time segment (layer) with occasional
  stochastic temporal progress without spatial transition
  (Fig.~\ref{fig:GPMN-random-walk}(b)). Each layer contains the
  original $G$ as is, while all the inter-layer edges are diagonal,
  i.e., connecting the same node from one time point to the next. The
  relative probability of such temporal progress can be adjusted by
  changing the average edge weight in $T$ (we call it $\langle E_T
  \rangle$) with reference to the average edge weight in $G$ (we call
  it $\langle E_G \rangle$). This can be a useful random walk model if
  time has well-defined segments (e.g., days, weeks, months) and many
  transitions are expected to occur within each segment before moving
  onto the next segment.
\item[Generalized product $\displaystyle G \GP_{\alpha, \beta} T$]
  This represents the combination of the above two random walk
  processes with their relative weights given by $\alpha$ and $\beta$
  (Fig.~\ref{fig:GPMN-random-walk}(c)). Therefore, together with the
  $\langle E_T \rangle / \langle E_G \rangle$ ratio, this model
  essentially has three adjustable global parameters. They
  collectively determine the relative likelihoods of (i) transitions
  within a single time segment (determined by $\alpha \langle E_G
  \rangle$), (ii) temporal progress without spatial transitions
  (determined by $\alpha \langle E_T \rangle$), and (iii) temporal
  progress with spatial transitions (determined by $\beta \langle E_G
  \rangle \langle E_T \rangle $).
\end{description}

Using this framework, interactions between nontrivial space and time
in the propagation processes can be written concisely, and the
spectral properties of the resulting GPMN can be obtained analytically
(or by approximation for Laplacian spectra when a non-Cartesian
product is used). To the best of our knowledge, such use of graph
product to represent interactions between nontrivial space and time is
novel in the literature.

For example, consider a random spreading process on a network
occurring in a cyclical time structure (e.g., a seasonal cycle in a
year), where $G$ is a typical transition probability matrix and $T$ is
a directed circular graph. Applying any of the above graph product
operators to these two factor networks produces a nonsimple GPMN model
of a weighted random walk process developing in both space and
time. The largest eigenvalue of the resulting adjacency matrix, an
indicator of the efficiency of spreading (and persistence), can still
be predicted using Eqs.~(\ref{eq:max-eigval}) or
(\ref{eq:lambdamax-formula}). If $\lambda^G_{\max}$ is assumed to be
constant, the efficiency of spreading is solely determined by the
adjacency spectrum of $T$.

\begin{figure}
\centering
\begin{tabular}{ll}
(a) & (b)\\
\includegraphics[width=0.49\textwidth]{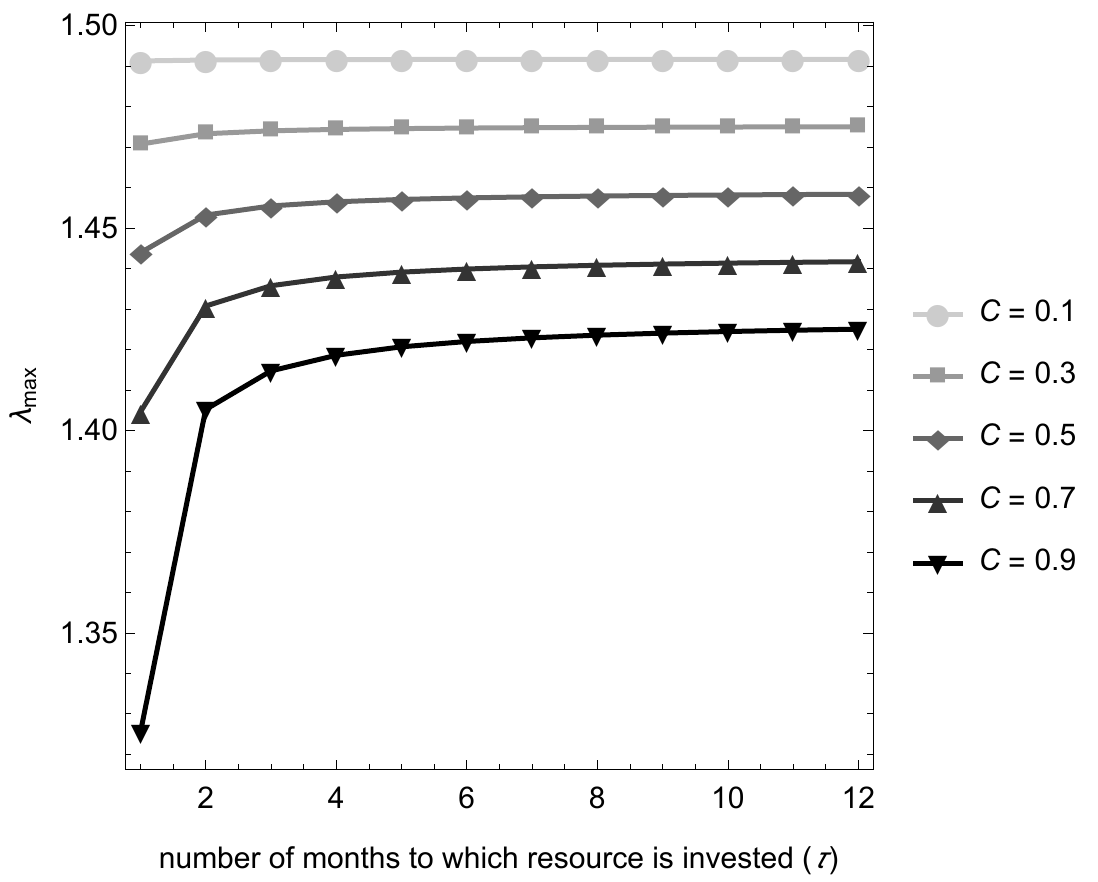} &
\includegraphics[width=0.49\textwidth]{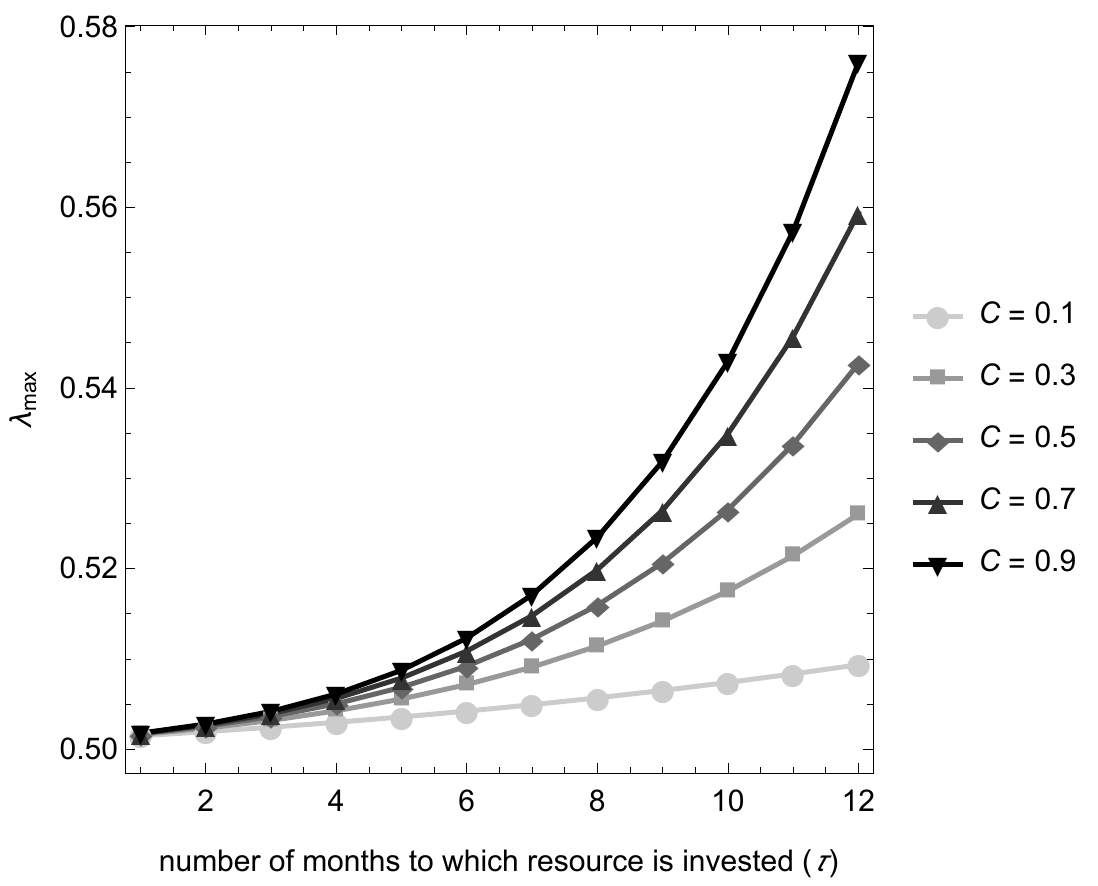}
\end{tabular}
\caption{Dependence of $\lambda_{\max}$ (spreading efficiency) on
  $\tau$ (number of months to which resource is invested),
  analytically obtained from adjacency spectra of factor networks
  using Eq.~(\ref{eq:lambdamax-formula}). The interactions between
  space and time was represented by a generalized product $G
  \GP_{0.5,0.5} T$, where $G$ is a $1000 \times 1000$ random
  transition probability matrix (fixed) and $T$ is a directed circular
  graph made of 12 nodes, representing 12 months = 1 year. Edge
  weights in $T$ were determined based on the investment strategy,
  described below. (a) Results from the epidemic control scenario, in
  which the default edge weight in $T$ was 1, and $\tau$ randomly
  selected edges in $T$ received a $C/\tau$ weight reduction each. (b)
  Results from the marketing scenario, in which the default edge
  weight in $T$ was $10^{-3}$, and $\tau$ randomly selected edges in
  $T$ received a $C/\tau$ weight increase each.}
\label{fig:epidemic-marketing}
\end{figure}

This model provides implications for temporal investment strategies
for epidemic control or marketing. In an epidemic control scenario, it
is reasonable to assume that $T$ has a large edge weight by default
and one needs to allocate a finite resource (denoted as $C$) to reduce
some (or all) of edge weights to suppress continuation of a disease
from one time point to another. In a marketing scenario, however, the
situation would be nearly the opposite; the default edge weight in $T$
is assumed to be near-zero and one needs to allocate the resource to
increase the chance of retention of product adoption over
time. Temporally heterogeneous investment strategies can be
efficiently represented by varying edge weights in $T$, and the
spreading efficiency (the largest eigenvalue of the entire adjacency
matrix of the GPMN) can be obtained efficiently using its spectral
properties. Figure \ref{fig:epidemic-marketing} presents a sample
result of such numerical computation for $G$ with 1000 nodes and $T$
with 12 nodes (time segments), clearly showing that temporally
concentrated investment was most effective for epidemic control, but
temporally distributed investment was necessary for marketing.

\section{Application III: Higher-Order Properties}
\label{sec:app3}

\begin{figure}
\centering
\begin{tabular}{ll}
(a) & (b)\\
\includegraphics[width=0.35\columnwidth]{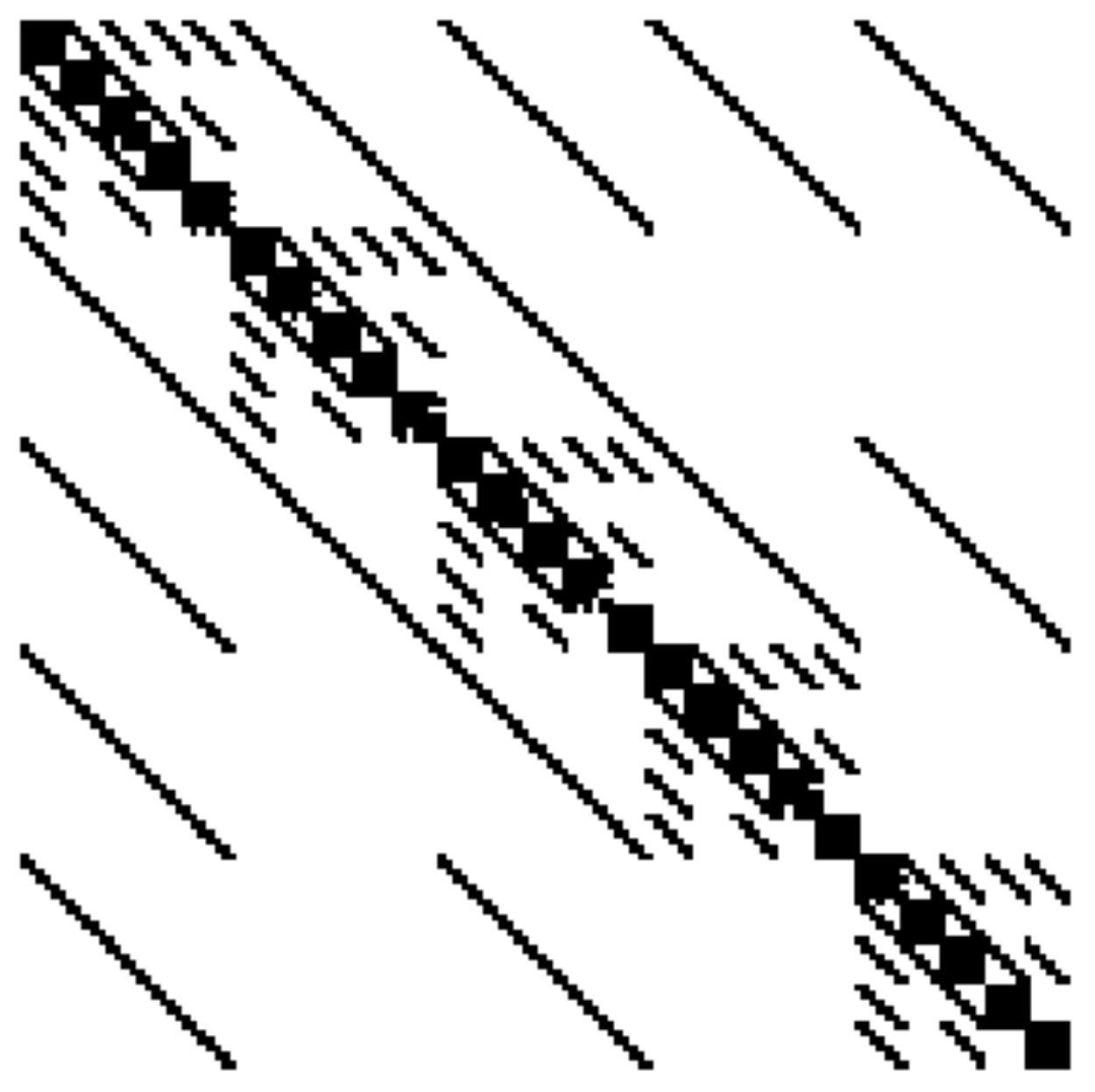} &
\includegraphics[width=0.35\columnwidth]{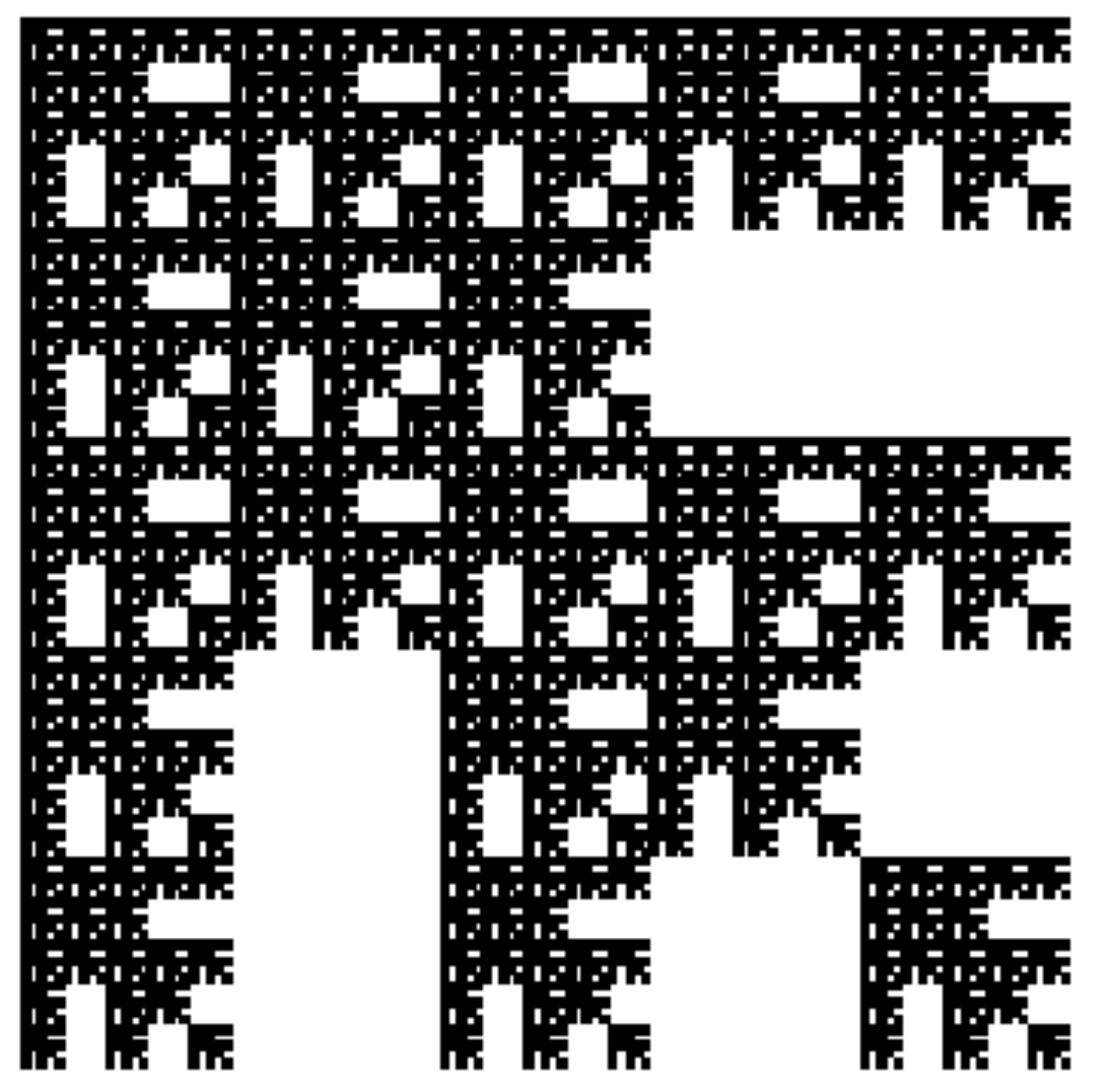} 
\end{tabular}
\caption{Examples of adjacency and Laplacian matrices of self-similar
  GPMNs of a higher-order power, showing fractal structure. The
  original network $G$ was a random graph with 5 nodes and 7
  edges. (a) Adjacency matrix of $G^5_{\CP}$. (b) Laplacian matrix of
  $G^5_{\SP}$. Non-zero elements are visualized with black pixels.}
\label{fig:fractal-matrices}
\end{figure}

Lastly, we discuss a rather different type of application of GPMNs:
{\em higher-order powers of a network.} Because graph product
operators we have considered are all associative, we can create GPMNs
by raising a network to the power of $n$ using any of the operators,
such as $G \CP G \CP G \CP \ldots \CP G = G^n_{\CP}$ (and similarly,
$G^n_{\DP}$, $G^n_{\SP}$, etc.). We call these GPMNs {\em self-similar
  GPMNs}, because their inter-layer structures are similar to their
intra-layer structures, and also because their adjacency and Laplacian
matrices asymptotically become fractal matrices
(Fig.~\ref{fig:fractal-matrices}) as the power $n$ increases.

\begin{figure}
\centering
\begin{tabular}{l}
(a)\\
\includegraphics[width=0.55\columnwidth]{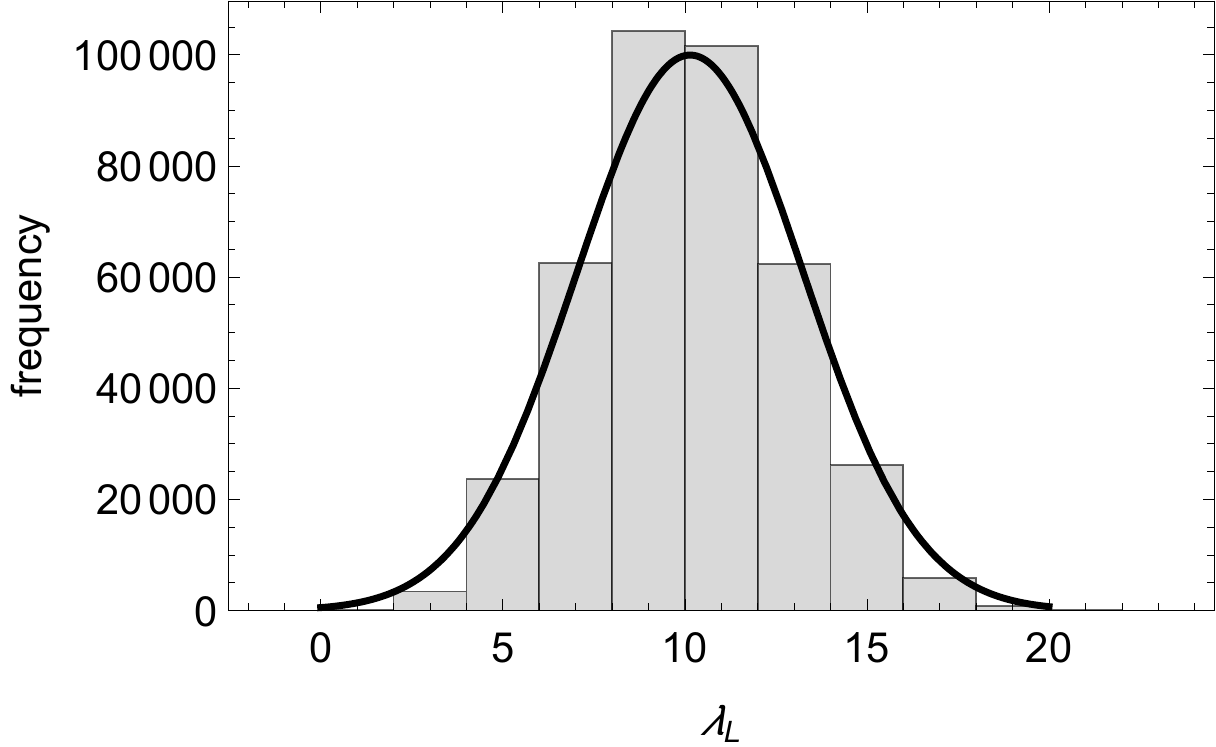}\\
~\\
(b)\\
\includegraphics[width=0.55\columnwidth]{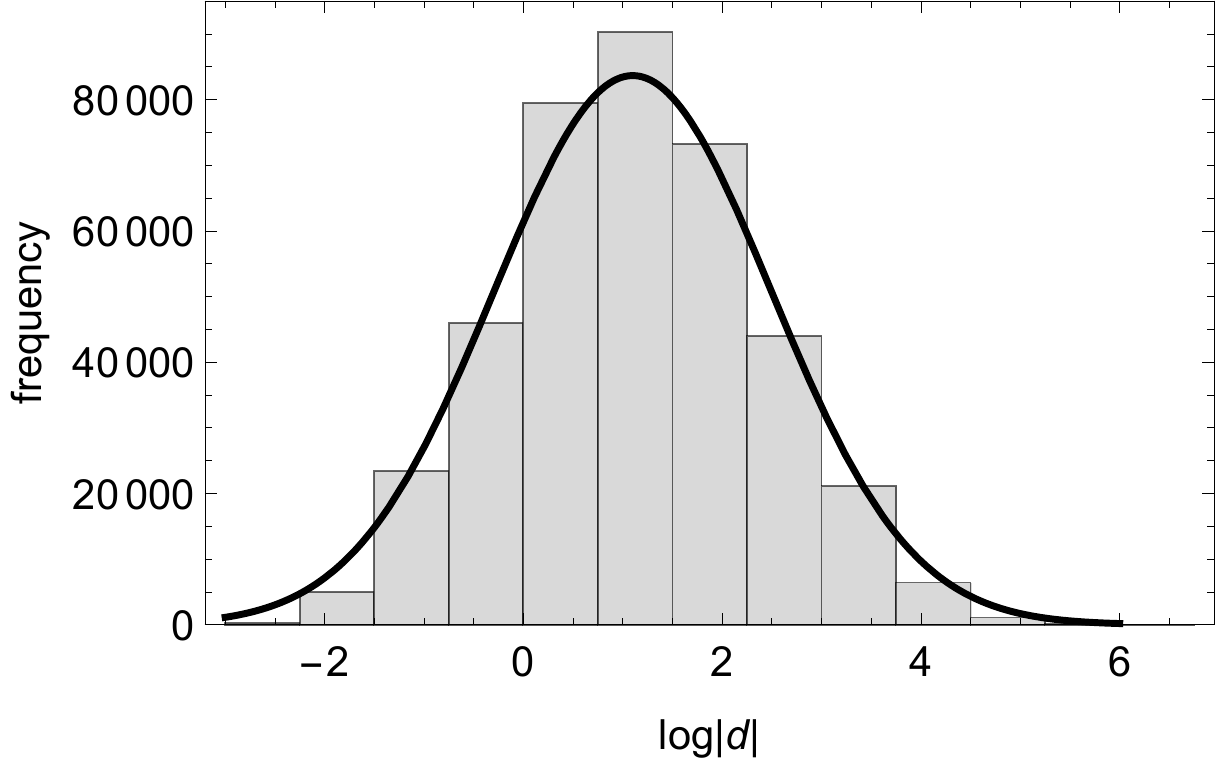}\\
~\\
(c)\\
\includegraphics[width=0.55\columnwidth]{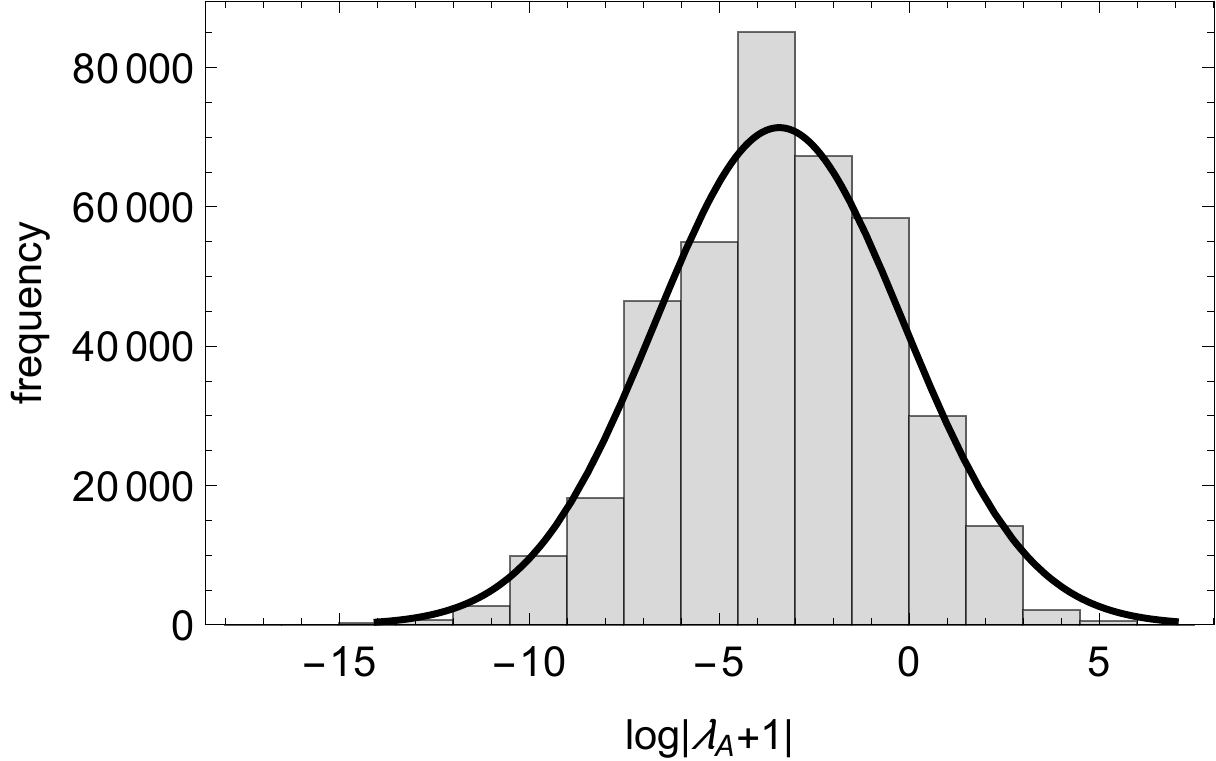}
\end{tabular}
\caption{Comparison of spectral distributions of self-similar GPMNs
  (histograms) and analytical predictions (curves). In all cases, the
  original network $G$ was a random network with 5 nodes and 7 edges,
  and each edge was assigned a random weight. (a) Laplacian spectrum
  of $G^8_{\CP}$. (b) Strength spectrum of $G^8_{\DP}$. (c) Adjacency
  spectrum of $G^8_{\SP}$.}
\label{fig:SS-asymptotic-spectra}
\end{figure}

We can analytically predict spectral distributions of self-similar
GPMNs for certain cases, as follows:
\begin{description}
\item[$G^n_{\CP} (\alpha = 1, \beta = 0)$] In this case, all the
  degree/strength, adjacency, and Laplacian spectra will be given by
  the sums of $n$ sample values independently selected (repetition
  allowed) from the original spectrum of $G$, to which the central
  limit theorem applies. Therefore, with $n \to \infty$, the
  asymptotic spectral distribution of $G^n_{\CP}$ will approach a
  normal distribution with mean $\langle x \rangle n$ and
  standard deviation $\sigma_x \sqrt{n}$, where $\langle x \rangle$ and $\sigma_x$
  are the mean and the standard deviation of the original spectrum,
  respectively. The smallest and largest values will also scale
  linearly with $n$.
\item[$G^n_{\DP} (\alpha = 0, \beta = 1)$] In this case, the
  degree/strength and adjacency spectra will be given by the products
  of $n$ sample values independently selected (repetition allowed)
  from the original spectrum of $G$. The adjacency spectrum may
  contain negative or even complex numbers (if $G$ is
  directed). However, if we take the logarithms of their absolute
  values, this again becomes the sums of $n$ sample values, to which
  the central limit theorem still applies. Therefore, with $n \to
  \infty$, the asymptotic distribution of absolute values of the
  degree/strength and adjacency spectra of $G^n_{\DP}$ will approach a
  log-normal distribution (with mean $\langle \log |x| \rangle n$ and
  standard deviation $\sigma_{\log |x|} \sqrt{n}$ in log space). The smallest and
  largest absolute values will scale exponentially with $n$.
\item[$G^n_{\SP} (\alpha = 1, \beta = 1)$] In this case, the
  degree/strength and adjacency spectra will be given by the same
  formula $\left( x_i + x_j + x_i x_j \right) = \left( (x_i + 1) (x_j
  + 1) - 1 \right)$ $\forall i, j$. The right hand side contains
  convenient $+1$ and $-1$, which indicates that distributions
  obtained by elevating all values in these spectra by 1 will behave
  exactly the same way as the degree/strength and adjacency spectra of
  $G^n_{\DP}$ discussed right above. Therefore, with $n \to \infty$,
  the asymptotic distribution of absolute values of the
  degree/strength and adjacency spectra of $G^n_{\SP}$, {\em when
    elevated by 1}, will approach a log-normal distribution (with mean
  $\langle \log |x+1| \rangle n$ and standard deviation $\sigma_{\log |x+1|}
  \sqrt{n}$ in log space). The smallest and largest absolute values
  (after elevated by 1) will scale exponentially with $n$.
\end{description}
Figure \ref{fig:SS-asymptotic-spectra} presents some examples of
spectral distributions of self-similar GPMNs together with analytical
predictions described above. Analytical predictions showed a good fit
to the actual spectral distributions.

An illustrative example of application of self-similar GPMNs to
dynamical networks is to use self-similar DPMNs to represent
simultaneous Markov processes on a network. Specifically, the $n$-th
power of a state transition network (such as $G$ given in
Fig.~\ref{fig:GPMN-random-walk}), when raised by direct product, gives
a higher-order transition network for meta-states of a population of
$n$ independent yet distinguishable Markovian individuals. This can be
easily understood in that each of the nodes of this self-similar DPMN
is a composite of $n$ independent choices of states in the original
transition network, and that the state transition probabilities
between those composite nodes are given by the $n$-th power (by
Kronecker product) of the original transition probability matrix,
i.e., the transition probability between two composite states is a
simple product of $n$ independent transition probabilities between
original individual states. This represents $n$ identical,
independent, simultaneous Markovian processes taking place on a
network, using a higher-order algebraic notation.

The results given in this section indicate that it is possible to
analytically predict some of the dynamical properties of systems
represented in self-similar GPMNs, including the simultaneous Markov
processes described above. We can characterize the spectral density
function and the scaling behavior of the largest/smallest values (in
either original values or in their absolute values). In particular,
the dominant eigenvalue of a self-similar GPMN is most likely
determined solely by the dominant eigenvalues of the original
network. Meanwhile, the fact that the spectra will approach a normal
or log-normal distribution implies that the less dominant modes of the
network's collective states will behave more and more homogeneously as
the order increases.

Finally, self-similar GPMNs can also be used to predict spectral
properties of certain high-dimensional networks of mathematical
interest. For example, an $n$-dimensional hypercube can be considered
a self-similar CPMN of a complete graph made of two nodes, raised to
the $n$-th power using Cartesian product. We can easily predict that a
high-dimensional hypercube's adjacency and Laplacian spectra will
asymptotically become a normal distribution centered at the origin and
at $n$, respectively, because the adjacency and Laplacian spectra of
the original two-node graph have $0$ and $1$ as their means,
respectively. This agrees with the results recently reported elsewhere
\cite{florkowski2008spectral}.

\section{Conclusions}
\label{sec:conclusions}

We have discussed GPMNs and their spectral properties. With the two
extensions introduced in this paper, GPMNs form an interesting, useful
family of multilayer networks that can provide an efficient,
analytically tractable framework for describing certain classes of
complex networks. GPMNs show mathematically nice behaviors in a number
of aspects, which can facilitate analytical and computational
investigation of structure and dynamics of various multilayer
networks. We also have presented several examples of applications,
which, we hope, have collectively illustrated the effectiveness and
practical value of the GPMN framework.

As mentioned earlier in Section \ref{sec:intro}, it would be highly
unlikely that the structure and dynamics of a real-world network could
be perfectly captured within the GPMN framework. While this is
certainly a limitation of GPMNs, it also suggests different roles for
GPMNs to play in the science of complex networks. Specifically, with
their simplicity and mathematically nice behavior, GPMNs may serve as
an analytically tractable approximation model, with which researchers
can produce analytical predictions and/or systematic comparison and
testing of empirically observed properties of multilayer
networks. This is similar to what mean-field models have been offering
to high-dimensional dynamical systems analysis. We believe that GPMNs
can contribute to the
multilayer networks literature in a similar manner.

There are a number of directions of further research. The
approximation of Laplacian spectra of DP/SP/GGPMNs requires further
development in both mathematical justification and performance
improvement. The three areas of applications should be evaluated
through more testing and validation using real-world network
data. There are also much room for further theoretical development and
exploration. For example, the implications of spectral properties of
GPMNs for canonical dynamical network models (e.g., diffusion,
synchronization, opinion formation, etc.) deserves more
elaboration. Another intriguing area of investigation would be the use
of nonlinear time structure, such as non-deterministic (branching)
flow of time, as a factor network in the spatio-temporal interaction
modeling, to explore other applications of GPMNs in different fields
of physics and computer science. Possible connections between
large-scale stochastic dynamics and GPMNs may also be worth further
investigation.

\section*{Funding}

This work was supported by the Visitors Program of the Max Planck Institute for the Physics of Complex Systems.

\section*{Acknowledgments}

The author thanks Lucas Wetzel for reviewing a draft version of this
work and providing helpful comments.

\appendix

\section*{Appendix: Spectral Relationships Between GGPMNs and Their Factor Networks}

Here we show that the strength (degree) and adjacency spectra of
GGPMNs still maintain exact algebraic relationships with those of
their factor networks, as follows:
\begin{description}
\item[In-strength spectra (= in-strength sequence)] With $\left( d^G_i \right)$ and
  $\left( d^H_j \right)$ being the node strengths of factor networks $G$ and $H$,
  respectively:
\begin{align}
\left(d^{G \GP_{\alpha, \beta} H}\right)
&= \left( A_G \KG_{\alpha, \beta} A_H \right) \bone_{|V_G||V_H|} 
= \left( \alpha A_G \KS A_H + \beta A_G \KP A_H \right) \left(\bone_{|V_G|} \KP \bone_{|V_H|} \right)\\
&= \alpha \left( A_G \KP I_{|V_H|} \right) \left(\bone_{|V_G|} \KP \bone_{|V_H|} \right) + \alpha \left( I_{|V_G|} \KP A_H \right) \left(\bone_{|V_G|} \KP \bone_{|V_H|} \right) + \beta \left( A_G \KP A_H \right) \left(\bone_{|V_G|} \KP \bone_{|V_H|} \right)\\
&= \alpha \left( d^G \right) \KP \bone_{|V_H|} + \alpha \bone_{|V_G|} \KP \left( d^H \right) 
+ \beta \left( d^G \right) \KP \left( d^H \right) \\
&= \underline{\left( \alpha d^G_i + \alpha d^H_j + \beta d^G_i d^H_j \right)} \quad \quad \forall i, j
\end{align}
Out-strength spectra can be obtained similarly by the same formula.
\item[Adjacency spectra] With $\left( \lambda^G_i, v^G_i \right)$ and $\left( \lambda^H_j,
v^H_j \right)$ being eigenvalues and eigenvectors of $A_G$ and $A_H$,
respectively:
\begin{align}
A_{G \GP_{\alpha, \beta} H} \left( v^G_i \KP v^H_j \right) &=
\left( A_G \KG_{\alpha, \beta} A_H \right) \left( v^G_i \KP v^H_j \right) = 
\left( \alpha A_G \KS A_H + \beta A_G \KP A_H \right) \left( v^G_i \KP v^H_j \right)\\
&= \alpha \left( A_G \KP I_{|V_H|} \right) \left( v^G_i \KP v^H_j \right) + \alpha \left( I_{|V_G|} \KP A_H \right) \left( v^G_i \KP v^H_j \right) + \beta \left( A_G \KP A_H \right) \left( v^G_i \KP v^H_j \right)\\
&= \alpha \left( \lambda^G_i v^G_i \right) \KP v^H_j + \alpha v^G_i \KP \left( \lambda^H_j v^H_j \right)
+ \beta \left( \lambda^G_i v^G_i \right) \KP \left( \lambda^H_j v^H_j \right)\\
&= \underline{\left( \alpha \lambda^G_i + \alpha \lambda^H_j + \beta \lambda^G_i \lambda^H_j \right)} \left( v^G_i \KP v^H_j \right)
\quad \quad \forall i, j
\end{align}
\end{description}

Moreover, we present that the Laplacian spectra of GGPMNs can also be
approximated using the strength (degree) and Laplacian spectra of
their factor networks, as follows:
\begin{description}
\item[Approximated Laplacian spectra] With $\left( \mu^G_i, w^G_i
  \right)$ and $\left( \mu^H_j, w^H_j \right)$ being eigenvalues and
  eigenvectors of in- or out-strength Laplacian matrices of $G$
  ($L_G$) and $H$ ($L_H$), respectively, and $\left( w^G_i \KP w^H_j
  \right)$ being used as pseudo-eigenvectors of $L_{G \GP_{\alpha,
      \beta} H}$ \cite{sayama2016estimation}:
\begin{align}
L_{G \GP_{\alpha, \beta} H} \left( w^G_i \KP w^H_j \right) &=
\left(D_{G \GP_{\alpha, \beta} H} - A_{G \GP_{\alpha, \beta} H} \right) \left( w^G_i \KP w^H_j \right)\\
&= \left( \alpha D_{G \CP H} + \beta D_{G \DP H} - \alpha A_{G \CP H} - \beta A_{G \DP H} \right) \left( w^G_i \KP w^H_j \right)\\
&= \left( \alpha D_G \KP I_{|V_H|} + \alpha I_{|V_G|} \KP D_H + \beta D_G \KP D_H 
 - \alpha A_G \KP I_{|V_H|} - \alpha I_{|V_G|} \KP A_H - \beta A_G \KP A_H \right) \nonumber\\
& \quad \quad \left( w^G_i \KP w^H_j \right)\\
&= \left[ \alpha L_G \KP I_{|V_H|} + \alpha I_{|V_G|} \KP L_H + \beta D_G \KP D_H - \beta \left(D_G - L_G \right) \KP \left(D_H - L_H \right) \right] \left( w^G_i \KP w^H_j \right)\\
&= \left( \alpha L_G \KP I_{|V_H|} + \alpha I_{|V_G|} \KP L_H + \beta L_G \KP D_H + \beta D_G \KP L_H - \beta L_G \KP L_H \right) \left( w^G_i \KP w^H_j \right)\\
&= \alpha \left(\mu^G_i w^G_i \right) \KP w^H_j + \alpha w^G_i \KP \left( \mu^H_j w^H_j \right)+ \beta \left(\mu^G_i w^G_i \right) \KP \left( D_H w^H_j \right)+ \beta \left(D_G w^G_i \right) \KP \left( \mu^H_j w^H_j \right) \nonumber \\
& \quad \quad - \beta \left(\mu^G_i w^G_i \right) \KP \left( \mu^H_j w^H_j \right)\\
&\approx \underline{\left( \alpha \mu^G_i + \alpha \mu^H_j + \beta \mu^G_i d^H_j + \beta d^G_i \mu^H_j - \beta \mu^G_i \mu^H_j \right)} \left( w^G_i \KP w^H_j \right) \quad \quad \forall i, j\\
& \quad \text{(with heuristic assumptions $D_G w^G_i \approx d^G_i w^G_i$ and $D_H w^H_j \approx d^H_j w^H_j$ \cite{sayama2016estimation})} \nonumber
\end{align}
\end{description}
We have conducted similar numerical experiments as in Section
\ref{sec:nonsimple} to confirm that sorting eigenvalues of factor
networks' Laplacian matrices by their real parts in an ascending order
is still most effective for approximation of GGPMNs' Laplacian spectra
(results not shown, as they were quite similar to
Fig.~\ref{fig:laplacian-approximation-nonsimple-sortings}).

Note that these spectral relationships described in this appendix hold
for both simple and nonsimple GGPMNs.

\bibliographystyle{unsrt}
\bibliography{sayama}

\end{document}